# Three-stage Collapse of the Long Gamma-Ray Burst from GRB 160625B Prompt Multiwavelength Observations

V. M. Lipunov[1,2], V. A. Sadovnichy[3], M. I. Panasyuk[1,4], I. V. Yashin[4], S. I. Svertilov[1,4], S. G. Simakov[2], D. Svinkin[5], E. Gorbovskoy[2], G. V. Lipunova[2], V. G. Kornilov[1,2], D. Frederiks[5], V. Topolev[1,2], R. Rebolo[6], M. Serra[6], N. Tiurina[2], E. Minkina[1,2], V. V. Bogomolov[1,4], A. V. Bogomolov[4], A. F. Iyudin[4], A. Chasovnikov[1,2], A. Gabovich[8], A. Tsvetkova[5], N. M. Budnev[7], O. A. Gress[2,7], G. Antipov[2], I. Gorbunov[2], D. Vlasenko[1,2], P. Balanutsa[2], R. Podesta[9], K. Zhirkov[1,2], A. Kuznetsov[2], V. Vladimirov[2], F. Podesta[9], C. Francile[9], Yu. Sergienko[8], A. Tlatov[10], O. Ershova[7], D. Cheryasov[2], V. Yurkov[8], and A. V. Krylov[2]

[1] Lomonosov Moscow State University, Physics Department, Vorobievy Hills, 1, Moscow 119991, Russia; lipunov@sai.msu.ru
[2] Lomonosov MSU, SAI, Universitetsky, 13, Moscow 119234, Russia
[3] Lomonosov Moscow State University, GSP-1, Vorobievy Hills, 1, Moscow, 119991, Russia
[4] Lomonosov Moscow State University, Skobeltsyn Institute of Nuclear Physics (SINP MSU), Vorobievy Hills, 1, Moscow, 119991, Russia
[5] Ioffe Institute, 26 Politekhnicheskaya, St Petersburg 194021, Russia
[6] Instituto de Astrofísica de Canarias, Lactea, E-38205, La Laguna, Tenerife, Spain
[7] Irkutsk State University, 20, Gagarin, Irkutsk 664003, Russia
[8] Blagoveschensk State Pedagogical University, Lenin, 104, Amur, Blagoveschensk 675000, Russia
[9] Universidad Nacional de San Juan, Observatorio Astronomico Felix Aguilar Benavidez 8175(O), 5413 San-Juan Argentina
[10] Kislovodsk Solar Station Pulkovo Observatory, Gagarin 100, Kislovodsk 357700, Russia

*Received 2021 September 3; revised 2022 May 26; accepted 2022 September 16; published 2023 February 7*

## Abstract

This article presents the early results of synchronous multiwavelength observations of one of the brightest gamma-ray bursts (GRBs) GRB 160625B with the detailed continuous fast optical photometry of its optical counterpart obtained by MASTER and with hard X-ray and gamma-ray emission, obtained by the Lomonosov and Konus-Wind spacecraft. The detailed photometry led us to detect the quasi-periodical emission components in the intrinsic optical emission. As a result of our analysis of synchronous multiwavelength observations, we propose a three-stage collapse scenario for this long and bright GRB. We suggest that quasiperiodic fluctuations may be associated with forced precession of a self-gravitating rapidly rotating superdense body (spinar), whose evolution is determined by a powerful magnetic field. The spinar's mass allows it to collapse into a black hole at the end of evolution.

*Unified Astronomy Thesaurus concepts:* Gamma-ray bursts (629); Black holes (162); Neutron stars (1108)

## 1. Introduction

More than 20 yr of optically localized long gamma-ray burst (GRB) studies have made it possible to clarify in general the main physical processes at the base of this phenomenon. However, a number of important questions remain unclarified. We believe that a long GRB occurs as a result of the collapse of the rapidly rotating core of a massive star. The fast rotation of the core slows down the collapse and extends the time available to produce electromagnetic radiation.

Two scenarios are possible here. In the first scenario (MacFadyen & Woosley 2001), a black hole first forms, and then the fallback of the supernova envelope, which has a supercritical torque, forms a heavy superdense disk. Due to the generation of magnetic fields by this disk, axial jets with a large gamma factor are generated, which we observe. In another scenario, a rapidly rotating magnetized object, a spinar, is first formed, which is slowly compressed due to the dissipation of the rotational moment (Lipunov & Gorbovskoy 2007). In this case, a jet with a Poynting–Umov energy flow is formed along the rotation axis. The operating time of the central GRB engine changes depending on the dissipation rate. In general, both scenarios require a fairly large torque in the collapsing stellar core. And here the authors are impressed by a binary scenario in which fast rotation occurs due to the tidal influence of the second component in a very close binary system (Tutukov & Cherepashchuk 2016). In this scenario, the centrifugal barrier is a consequence of the large torque acquired as a result of the natural evolution of the binary system. In the case of GRB 160625B our attention was attracted by the quasiperiodic brightness fluctuations during the time of the central engine operation, and below we try to interpret them as a consequence of a spinar paradigm at work (Lipunova & Lipunov 1998; Lipunov & Gorbovskoy 2007, 2008; Lipunova et al. 2009).

## 2. Observations

Due to its outstanding brightness, GRB 160625B was observed by a large number of space and ground telescopes in a wide range of electromagnetic waves from gamma-ray to radio (e.g., Burns 2016; Barthelmy et al. 2016; Dirirsa et al. 2016; Evans 2016; Kuroda et al. 2016; Lipunov et al. 2016a; Melandri et al. 2016; Oates 2016; Troja et al. 2016; Topolev & Lipunov 2021). Many papers have already been published, including some by the authors of this work (Alexander et al. 2017; Troja et al. 2017; Ravasio et al. 2018, 2019; Zhang et al. 2018; Strausbaugh et al. 2019; Kangas et al. 2020, and others). Particularly, for the first time in the history of the study of GRBs, the polarization of its own optical radiation synchronous with the gamma was detected (Troja et al. 2017). This paper intends to present unpublished details of synchronous observations in the optical and gamma-ray ranges.

On 2016 June 25 at 22:40:16 UT, the Enrique Fermi Space Observatory (Meegan et al. 2009) detected the most powerful GRB in history, GRB 160625B, first as a short pulse (Fermi-GBM trigger 488587220; Burns 2016). At 22:43:24.82 Fermi-LAT triggered on a bright pulse from the same GRB (Dirirsa et al. 2016). At 22:51:16.03 Fermi-GBM was triggered again (trigger 488587880). The GBM light curve consists of multiple peaks over approximately 700 s, the first one being a 1 s long soft peak. The main peak, corresponding to the LAT trigger, was very hard and about 25 s long. The peak that triggered GBM for the second time was soft and about 30 s long (Burns 2016).

This long, extremely bright GRB 160625B also triggered detectors BDRG on board the Lomonosov Space Observatory of Moscow State University (Sadovnichii et al. 2017) and Konus-Wind at 22:40:19.875 UT (Svinkin et al. 2016) as well as the CALET Gamma-ray Burst Monitor (CGBM) at 22:40:15.49 (Nakahira et al. 2016). Swift-XRT detected an uncatalogued X-ray source at position R.A., decl. J2000: 20:34:23.25, +06:55:10.5 (Melandri et al. 2016), which was an X-ray counterpart of GRB 160625B.

The optical counterpart was discovered by RATIR (Troja et al. 2016) starting 8.53 hr after LAT triggered. The detection of a substantial ($8.3\% \pm 0.8\%$ from our most conservative estimation) variable linear polarization of the prompt optical flash that accompanied the extremely energetic and long prompt gamma-ray emission from GRB 160625B was discovered by MASTER (Gorbovskoy et al. 2016a; Lipunov et al. 2016a; Troja et al. 2017). MASTER started observations in polarization filters (Lipunov et al. 2010, 2019; Kornilov et al. 2012) 31 s after GBM notice time (57 s after GBM, i.e., 131 s before LAT trigger; Dirirsa et al. 2016) at 2016 June 25 22:41:13 UT (Gorbovskoy et al. 2016a), but published in a Gamma-Ray Coordinates Network (GCN) circular (Barthelmy 1998; Barthelmy et al. 1998) at 2016 June 28 14:05:38. The GCN publication was delayed by the installation of the new MASTER telescope in Argentina (MASTER-OAFA). MASTER measurements probed the structure of the magnetic field at an early development stage of the jet, closer to a central black hole, and show that the prompt-emission phase is produced via fast-cooling synchrotron radiation in a large-scale magnetic field that was advected from the black hole and distorted by dissipation processes within the jet (Troja et al. 2017). The optical data obtained by MASTER telescope robots (Lipunov et al. 2010, 2019) have the best temporal resolution with a minimum exposure time of 5 s. This resolution made it possible to suspect quasiperiodic variability in the optical range, which we try to associate with the dual nature of the long GRB.

## 2.1. MASTER Observation

The MASTER Global Robotic Net of Lomonosov Moscow State University in 2016 consisted of eight ground observatories with identical scientific equipment distributed all over Earth: MASTER-Amur, -Tunka, -Ural, -Kislovodsk, -Tavrida in Russia, -SAAO (South African Astronomical Observatory), -IAC (Tenerife, Spain, Teide Observatory of the Institute of Astrophysics of the Canary Islands), and -OAFA (Argentina, San Juan National University Astronomical Observatory named by Felix Aguilar), see Lipunov et al. (2010, 2019).

Identical equipment includes twin wide-field (MASTER-II) and very wide-field (MASTER-VWFC) optical channels (Kornilov et al. 2012), which allow us to follow a target 24 h per day in one photometric channel. Two of the same MASTER-VWF cameras are also mounted on the Lomonosov University space observatory (Sadovnichy et al. 2017; Lipunov et al. 2018a). Observations by MASTER-VWFC are unfiltered. MASTER-II includes 40 cm telescopes with a 4–8 deg$^2$ field of view (FOV; open and closed mode of observations). Our own photometer can simultaneously observe in two orthogonally oriented polarization filters (Kornilov et al. 2012) or in $BVRI$ filters and unfiltered. Mount has fast positioning with speed up to $30° \text{ s}^{-1}$. The key factors of full robotization include hardware control, weather control, smerides, central planner, automatic evening/morning calibration, and primary image reduction, astrometry and photometry, extraction of new optical sources and notification of the main station of them). Such features led us discover significant and variable linear polarization during the prompt optical flash of GRB 160625B (Troja et al. 2017) to discover GRB optical counterparts (Lipunov et al. 2016a, 2017d; Gorbovskoy et al. 2016a, 2016b; Sadovnichy et al. 2018; Laskar et al. 2019; Ershova et al. 2020) to discover smooth optical self-similar emission of GRB subclass (Lipunov et al. 2017d), create the most optical support to the GW150914 event (Abbott et al. 2016; Lipunov et al. 2017a, 2017b, Lipunov et al. 2018b), independently discover kilonova GW170817 (Abbott et al. 2017a, 2017b; Lipunov et al. 2017c), create the most optical support for the follow-up of a rare IceCube neutrino multiplet (IceCube Collaboration et al. 2017), discover V404 Cygni polarization variability (Lipunov et al. 2016b), make optical observations that revealed strong evidence for a high-energy neutrino progenitor—blazar TXS 0506+056 for IC 170922A (Lipunov et al. 2020), and discover more than 2000 optical transients of 10 different types and other (Lipunov et al. 2016c).

All MASTER telescopes are connected to the GCN directly and have a secondary link through the MASTER central server (in Lomonosov MSU), which distributes plans of observations taking into account current tasks. When a GRB alert comes and its altitude is more than 0° at a site, the MASTER telescope interrupts its current exposition and moves to the GRB coordinates during the CCD readout cycle. Apart from GRBs, gravitational wave, high-energy neutrino alerts follow-up observations, MASTER telescopes follow sky surveys on their own.

After a fainter second peak at T0+15.9 s of GRB 160625B, the optical counterpart flux declined steadily. During this phase, the MASTER-IAC telescope in Tenerife, Spain, observed the optical counterpart in two orthogonal polaroids simultaneously, starting at T0+95 s and ending at T0+360 s.

The MASTER-IAC robotic telescope pointed to GRB 160625B 26 s after Swift notice time (Gorbovskoy et al. 2016a), 31 s after Fermi-GBM notice time (57 s after GBM triggered, which is equal to 131 s before LAT triggered (Dirirsa et al. 2016) on 2016 June 25 at 22:41:13 UT, and having postponed current observations, began/covering the error box with optical cameras. Starting from this time we continuously observed the Swift X-ray error box (Melandri et al. 2016) with the MASTER-IAC wide-field camera with a 5 s exposition time without a filter and imaged several thousand frames.

Despite the significant errors in the primary coordinates of the GRB, the entire error box was covered by very wide-field

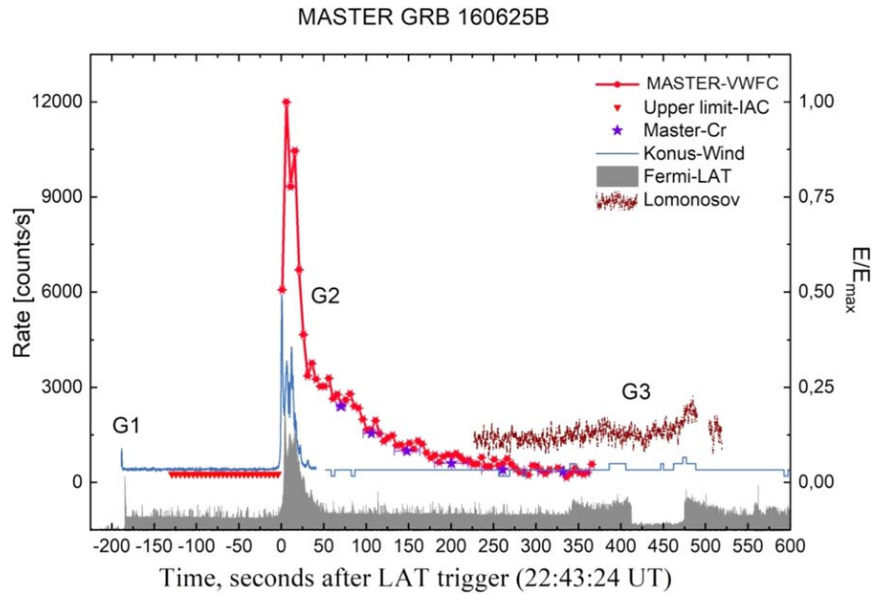

**Figure 1.** The red curve represents the observations of the very wide-field MASTER-IAC cameras, reduced to the proper time of the system (compression of the burst timescale relative to the observed one). The blue asterisks correspond to the MASTER-Tavrida measurements taken at a lower temporal resolution. The gray curve corresponds to the emission detected by Fermi-LAT.

cameras—MASTER-VWFC, the ground-based analog of the MASTER-SHOCK cameras installed on board the Lomonosov space observatory (Sadovnichii et al. 2017; Lipunov et al. 2018b; Park et al. 2018; Sadovnichy et al. 2018; Svertilov et al. 2018), which were designed for such cases.

MASTER-Tavrida and MASTER-IAC's very wide FOV cameras were already observing a GRB region when, 131 s after the first message, Fermi-LAT detected the main impulse of the GRB with a high coordinate accuracy of 0°.5. MASTER-Tavrida became a new site of the MASTER Global Network in 2016 and worked in a test mode. This telescope started alert observations (first frame exposition) 12 s after LAT notice time (66 s after trigger time) at 22:44:30 UT.

The optical counterpart was detected at 22:43:30 (+2.5 s of exposition) UT with $m_{OT} = 8^{m}.60$ and was becoming brighter up to a maximum of $7^{m}.86$ (see Figure 1 and tables). The prompt optical emission strongly correlated with the Konus-Wind gamma-ray light curve (Svinkin et al. 2016).

The MASTER-IAC robotic telescope started the observation of the error box 43 s after LAT notice time or 95 s after the trigger time on 2016 June 25 at 22:44:59 UT by the main MASTER-II telescope in two polarizations. The OT was $8^{m}.6$ at the moment.

As a result, MASTER not only registered the entire event of the GRB explosion in the optical range with the best time resolution (2.5 s), but for the first time detected the polarization of prompt optical emission from the GRB while the flash was still occurring (Troja et al. 2017). Photometry results for GRB 160626B are presented in the Appendix.

The GRB GRB 160625B turned out to be one of the most powerful cosmic explosions of this type, which appeared as a narrow jet of relativistic particles accelerated by the electromagnetic field of a rapidly rotating black hole at the other end of the universe forming before our eyes.

An analysis of the MASTER's polarization observations made it possible for the first time to detect the polarization of the GRB's intrinsic optical emission and directly showed that the muzzle of the most powerful space gun was formed by an ordered powerful magnetic field of a forming black hole (Troja et al. 2017).

This magnificent astrophysical experiment succeeded thanks to the interaction of scientists from several countries, who created unique robotic equipment to detect gamma rays, infrared radiation, and photons in the optical range born by the GRB event.

### 2.2. Lomonosov Observations

The GRB monitor (BDRG) on board the Lomonosov (Sadovnichii et al. 2017) observatory (hereafter BDRG/Lomonosov (Svertilov et al. 2018) was built for the early detection of GRBs in the 0.01–3.0 MeV energy range and for generation of triggers for those events. The BDRG consisted of three identical detector units connected to the electronic unit. The BDRG instrument detector units (blocks) were mounted on the spacecraft payload platform in such a way that their axes are oriented 90° to each other. Each detector has a cosine angular dependence for a sensitive area not shaded by satellite construction elements within ~60° of its axis. The monitor's central axis, relative to which the detector axes are inclined, is directed toward to the local zenith. Thus, the total FOV for all three detectors is about $2\pi$ sr; and one-quarter of this field, i.e., $\pi/2$ sr, is the value of a solid angle, within which limits the GRB position error can be estimated with sufficiently good accuracy through the comparison of all three detector outputs.

BDRG operates in two main observational modes: the monitor or continuous mode, and the burst mode. In the monitor mode all instrument outputs were recorded and stored continuously with time resolutions adjustable by commands from Earth. On the other hand, the burst mode was activated by onboard instrument triggers to record detailed information of each photon during the before-burst, burst, and after-burst time intervals. The BDRG trigger initiated the estimation of GRB positions and relayed the trigger not only to other GRB observation instruments on board the Lomonosov spacecraft, i.e., SHOK optical cameras (Lipunov et al. 2018b) and the Ultrafast Flash Observatory (Park et al. 2018), but to the ground telescopes as well through the GCN (Barthelmy 1998; Barthelmy et al. 1998) via the Global Star transmitter.

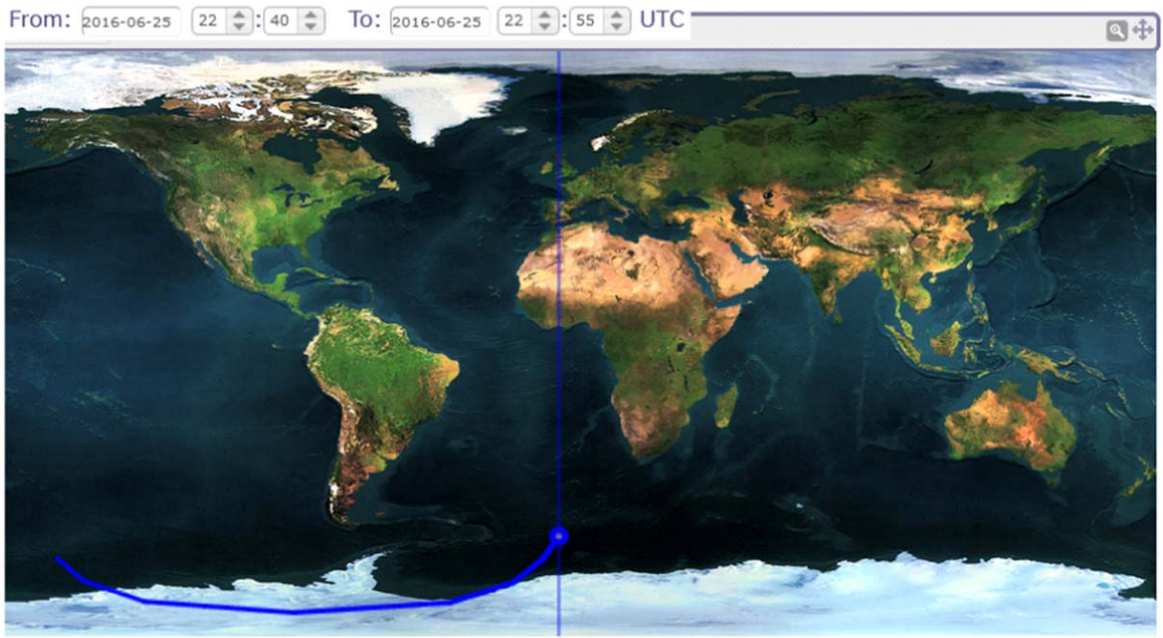

**Figure 2.** Projection of the Lomonosov satellite orbit (blue line) on a map of Earth. The blue spot corresponds to the GRB 160625B Swift trigger, i.e., 22:40:31 UTC.

Each BDRG detector unit consisted of a thin layer (0.3 cm) of NaI(Tl) crystals optically coupled to a considerably thicker layer (1.7 cm) of CsI(Tl) crystals situated underneath. The diameter of the detectors is 13 cm, and both layers are read by a single photomultiplier tube. Thus, the overall detector area is about 130 cm$^2$. The thickness of the NaI(Tl) layer is optimized for the soft part of the energy range, and the working ranges of the units are 0.01–0.5 MeV for the NaI(Tl) layer and 0.05–3 MeV for CsI (Tl). As such, the NaI(Tl) layer serves as the main detector for hard X-ray timing, while the CsI(Tl) operates as an active shield against background gamma rays. Additionally, the CsI(Tl) crystals can also detect gamma rays with energies up to a few megaelectronvolts. The difference in decay times for the NaI(Tl) (∼0.25 ms) and CsI(Tl) (∼2.0 ones) crystals permits the separation of light flashes in the scintillators through special electronic circuits that differentiate pulse shapes.

The information provided by the BDRG units consisted of a number of different categories for the data frames generated continuously (continuous mode) as well as irregularly by various triggers (burst mode). The continuous data stream included three types of frames corresponding to the instruments' monitoring, spectrum, and event. Monitoring frames provided count rates in eight energy channels for the NaI(Tl) and CsI(Tl) scintillator crystals for each of the BDRG detector units, while spectrum frames contain 724 channel spectra for NaI(Tl) and CsI(Tl), separately. Event frames provided the primary values for energy release within the NaI(Tl) and CsI (Tl) crystals, combined with time data for a fixed number of detected gamma quanta. Likewise, information about the main parameters for all GRB triggers was stored and transferred in the form of *trigger logs*. There are three trigger types categorized as *fast*, *slow*, and *super slow*, with characteristic times of 10 ms, 1 s, and 20 s, respectively. Corresponding to each trigger type, three data frame sequences for the monitoring, spectrum, and event were generated continuously in a manner similar to the continuous mode discussed above. A portion of data collected before the trigger was always included for all trigger types.

On 2016 June 25 near the GRB 160625B trigger, BDRG/Lomonosov operated in monitoring mode. The background environment at the time of the event was very complicated. The Lomonosov satellite was flying through the radiation belts (see Figure 2).

The count rate variations in the BDRG/Lomonosov gamma-quanta channels are illustrated by the curves in the upper panel of Figure 3, in which count rates of three BDRG1 NaI detector channels 10–35, 35–170, 170–650 keV shown with the time resolution of 0.1 s are plotted. The middle panel of Figure 3 shows data from the spectrometer SPI on board INTEGRAL obtained for the same time period (Mazaeva et al. 2016). The GRB was detected with the anticoincidence system (ACS), which provides integral counts of gamma quanta with energies of more than about 50 keV. For a detailed description of SPI/INTEGRAL's GRB detection capabilities see von Kienlin (2003).

In the top panel of Figure 4 a clear count rate increase in the time interval of about −150 to 350 s relative to the Swift trigger time (−300 to 200 s relative to the LAT trigger time) corresponds to the satellite crossing outskirts of the external radiation belt. After that, the satellite flew into the region of the south polar cap, where the background is smaller, and after it began to cross the outer belt again and the background began to increase. Thus, due to such background variations, the main pulse (G2 in Figure 1) as well as the precursor (G1 in Figure 1) could not be observed by the Lomonosov satellite detectors. Only the *last tail* (G3 in Figure 1) could be observed. However, during the corresponding time interval (of about 350–550 s from the LAT trigger time) count variations in BDRG/Lomonosov were in reality a combination of GRB counts and variations caused by unstable electron fluxes in the polar cap. The last component can provide significant input to the gamma-ray channels of the instrument due to electron bremsstrahlung. Thus, to obtain the real GRB light curve it is necessary to clean the detector outputs from electron background variations.

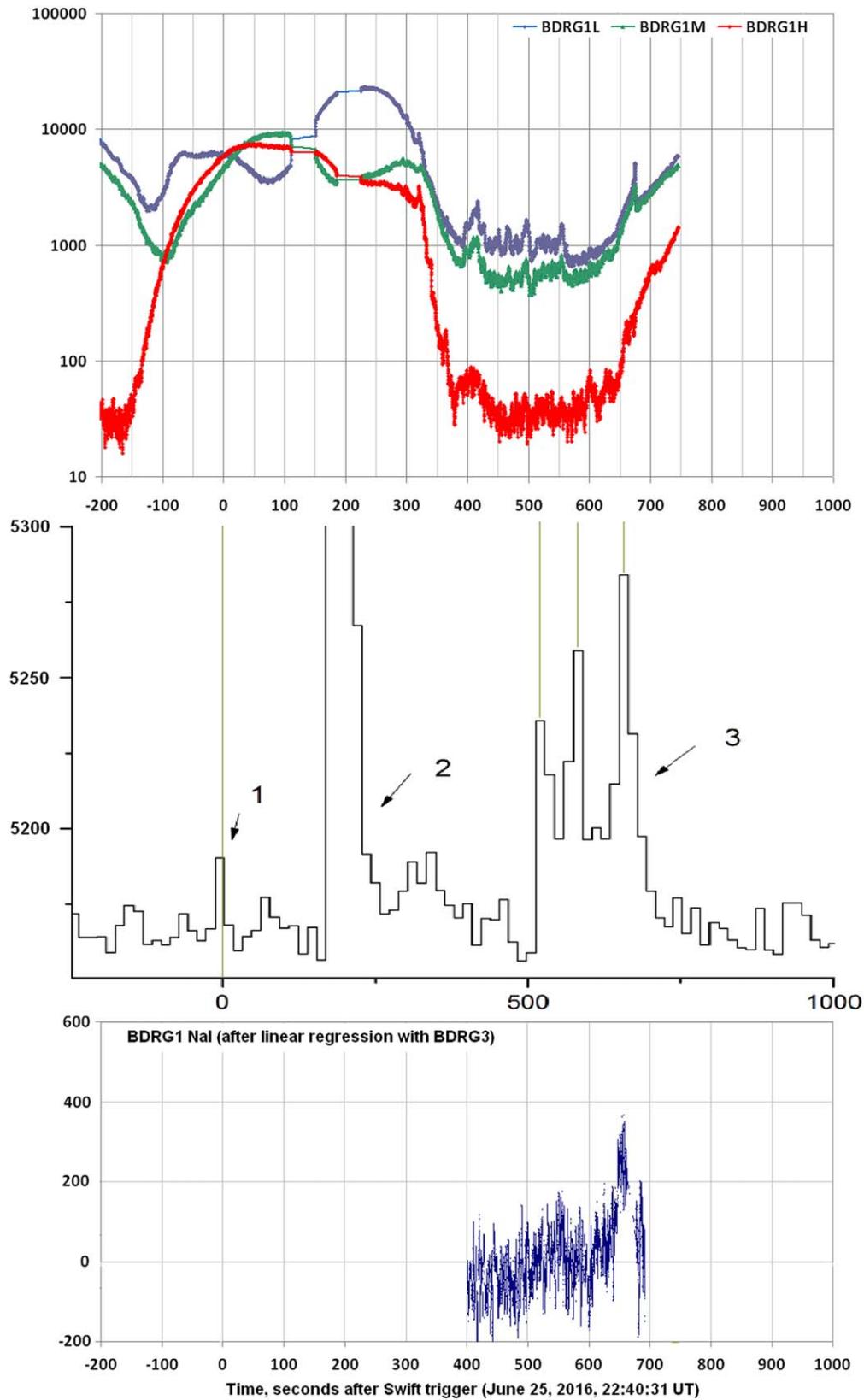

**Figure 3.** GRB 160626B gamma-ray light curves. The top panel is BDRG1 (NaI detector) counts vs. time in the 10–35 keV (BDRG1L), 35–170 keV (BDRG1M), and 170–650 keV (BDRG1H) ranges, the middle panel illustrates the SPI-ACS/INTEGRAL counting rate, and the bottom panel represents the light curve in the BDRG1M channel cleared from electron background variations.

To realize this procedure we used outputs from different BDRG detectors. Because of different BDRG detector orientations, the given GRB source was observed by separate detectors at different angles. In the case of GRB 160625B, only the BDRG1 detector unit was illuminated, and the angle between the detector axis and the direction toward the GRB source was about 56°, while for BDRG2 and BDGR3 these angles were about 136° and 116°, respectively. This means that the GRB source was out of the FOV of both of the latter detector units. On the other hand, since the counting rate of the irregular variations observed in the polar cap regions are mainly due to the quasi-trapped electron fluxes having an anisotropic but rather wide pitch-angle distribution, they will exhibit similar temporal behavior in separate, although differently oriented detectors. This allows us to use regression analysis of detector unit count rates obtained for two detectors during the time interval of a GRB observation to estimate regression coefficients, which can then be used for rejection of the part of the counting rate variations caused by electron fluxes. Because during the time of the GRB 160625B observation BDRG2 unit was switched off, for regression analysis only the BDRG1 and BDRG3 unit outputs were used. To be exact, we selected the time interval of 200–450 s after the LAT trigger (22:43:24 UTC) to estimate regression coefficients, which correspond to the time between the G2 and G3 events, when expected input to detectors counting rates from GRB 160625B was negligible.

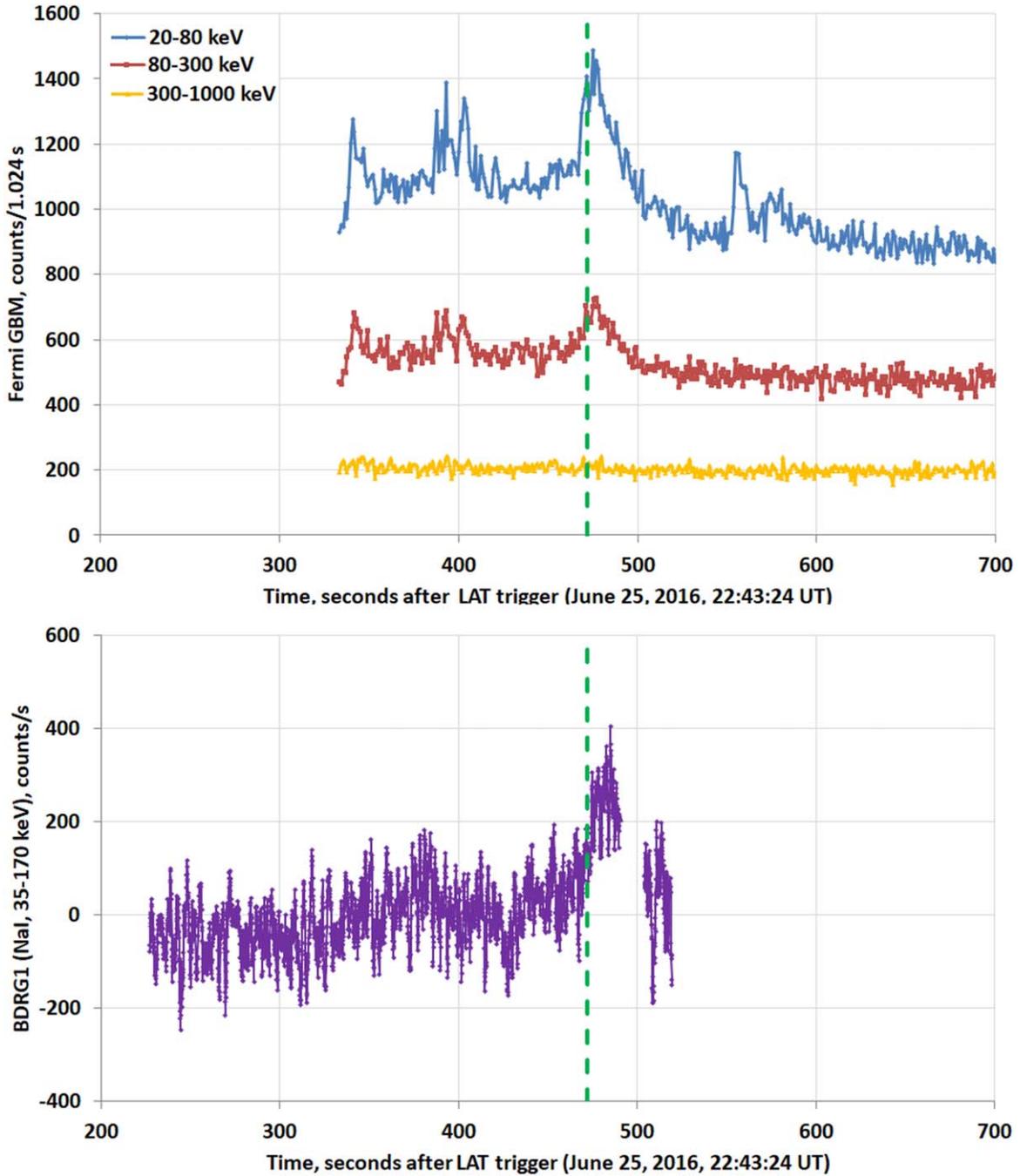

**Figure 4.** The top panel represents the GBM/Fermi counting rate vs. time in the 20–80, 80–300, and 300–1000 keV energy ranges, and the bottom panel represents the light curve from the BDRG1 (NaI detector) 35–170 keV channel cleaned from electron flux background variations, i.e., the zero counts correspond to the mean background level. The green dashed line represents the GBM/Fermi trigger (22:51:16 UT).

We use three types of regression models. In the first one (Model A), it was assumed that there is a linear correlation between counting rates $N_1$, $N_3$ in the same energy channels of the BDRG1 and BDRG3 NaI(Tl) detectors:

$$N_1 = kN_3 + B.$$

In the second model (Model B), the linear correlation between the counting rate in the given BDRG1 NaI(Tl) detector channel $N_1$ and counting rates in three BDRG3 NaI(Tl) energy channels, i.e., $L_3$ (low, 10–35 keV), $M_3$ (middle, 35–170 keV), and $H_3$ (high, 170–650 keV) was assumed:

$$N_1 = k_L L_3 + k_M M_3 + k_H H_3 + B.$$

And in the third model (Model C), it was assumed that the quadratic connection between counting rates in the given BDRG1 and BDRG3 NaI(Tl) detector channels is

$$N_1 = k_1 N_3^2 + k_2 N_3 + B.$$

The results of regression analysis with the use of all the mentioned models indicated that linear correlation, i.e., Model A is enough for the elimination of background variations (Models B and C did not provide a qualitatively better result). The best cleaning from electron background variations was obtained for the energy channel 35–170 keV, and the corresponding light curve is presented in the bottom panel of Figure 3 and 4.

The more detailed light curve also cleaned from electron variations is shown in Figure 4, where for comparison light curves obtained from GBM/Fermi instrument (A. von Kienlin, private communication) data, see https://heasarc.gsfc.nasa.gov/FTP/fermi/data/gbm/triggers/2016/bn160625952/.

The presented light curves point to a significant increase of hard X-ray and soft gamma-ray flux near 650 s after the BAT/Swift trigger or about 450 s after the GBM/Fermi trigger clearly, which is evidence of the central engine operation 10 minutes after the main explosion. The light curve recorded by the BDRG/Lomonosov instrument with a time resolution of 0.1 s demonstrates fine structure as a number of pulses, the most intensive at 22:51:16 UTC and less intense and wider at about 100 s before. This time structure is confirmed by GBM/Fermi data. The intensive pulse at 22:51:16 UTC as well as the preceding observed clearly at the energy range of 3.4–44 keV, while from the BDRG/Lomonosov data they could be observed in an energy interval of 25 keV up to 170 keV. The event is observed at about 140 s after the first burst peak, which implies that the GRB central engine continues its operation for a rather long time following burst inception. The fine structure observed on the GRB light curve dozens of seconds after the beginning of the central engine operation may be caused by collisions of relativistic shells propagated in beams with different Lorentz factors.

Moreover, the mean flux was estimated in the range of 35–170 keV for a time interval of 2–17 s from the Fermi trigger at 22:51:16 UT, i.e., for the count rate increase corresponding to the third episode of GRB source activity. For this, we used the modeling of the BDRG instrument response to the detection of gamma quanta falling on the detector under a given angle of about 60°, which was estimated from a known BDRG1 axis orientation and GRB source location. With this, we took into account the quanta detection in the NaI(Tl) crystal as well as an absorption in the instrument case, i.e., for the direction on the GRB source it was about 0.1 cm in Al equivalence. It was calculated that an effective area for the full absorption in NaI (Tl) was about 60 cm$^2$ and changed weakly in the energy range of 35–170 keV.

For the given time interval and energy range about 2400 quanta were detected. This corresponds to the spectral flux density 0.3 cm$^{-2}$ s$^{-1}$ keV$^{-1}$. It is necessary to note that statistical errors as well as uncertainties due to the real dependence of detection efficiency from the gamma-quantum energies are much less than the uncertainty associated with the regression model of background subtraction. This error was estimated empirically by the variation of regression coefficients, and its value is $(-0.2, 0.4)$ cm$^{-2}$ s$^{-1}$ keV$^{-1}$.

To recalculate the obtained spectral flux density into the energy fluence we need to know the real energy spectrum. Because it was difficult to obtain such spectrum from one energy range, we used the spectral parameters of the CPL models presented by Lü et al. (2017) for the discussed time interval. The normalizing factor in this model was determined from the condition that the count rates expected from the model should be close to the measured ones. Taking into account the above error interval presented we obtained the energy fluence kiloelectronvolt per square centimeter per second in the range of 35–170 keV for the time interval of 2–17 s from the Fermi-GBM trigger in a third episode of GRB 160625B source activity.

### 2.3. Konus-Wind Observations

GRB 160625B triggered the Konus-Wind GRB spectrometer (KW; Aptekar et al. 1995) at $T0 = 81{,}619.875$ s (22:40:19.875 UT, Svinkin et al. 2016). The burst was detected by the S2 detector, which observes the northern ecliptic hemisphere; the incident angle was 65°.2. The propagation delay from Earth to Wind is 3.356 s for this GRB; correcting for this factor, the KW trigger time corresponds to the Earth-crossing time 81,616.519 s (22:40:16.519 UT).

#### 2.3.1. Time History

In the KW trigger mode, count rates are recorded in three energy bands: 17–70 keV (B1), 70–300 keV (B2), and 300–1170 keV (B3). The record starts at $T0-0.512$ s and continues to $T0+29.376$ s with an accumulation time varying from 2–256 ms. Waiting-mode count rate data are available up to $T0+250$ s in the same energy bands with a coarse temporal resolution of 2.944 s. A source activity after the time interval $T0+250$ s may be traced in the housekeeping mode with a temporal resolution of 3.68 s for the 70–300 keV energy range.

The prompt-emission light curve (Figure 5) can be divided into three episodes. It starts at ∼$T0-0.3$ s, with a short, spectrally soft initial pulse (precursor), which has a duration of about 1 s. The precursor is followed, starting at ∼$T0+80$ s, by the main, extremely bright and spectrally hard emission episode lasting about 40 s. The final episode observed by KW in the housekeeping mode starts at ∼$T0+530$ s and has a duration of about 150 s. The total burst duration is about 680 s.

The first episode was only localized by Fermi (GBM) with a position uncertainty of about 2° (statistical only). The difference between the arrival time of the gamma-ray signals at Fermi-GBM and Konus-Wind provides additional significant constraints on the gamma-ray localization of the episode (Zhang et al. 2018). The triangulation of the first episode is consistent with the source position determined by Swift-XRT

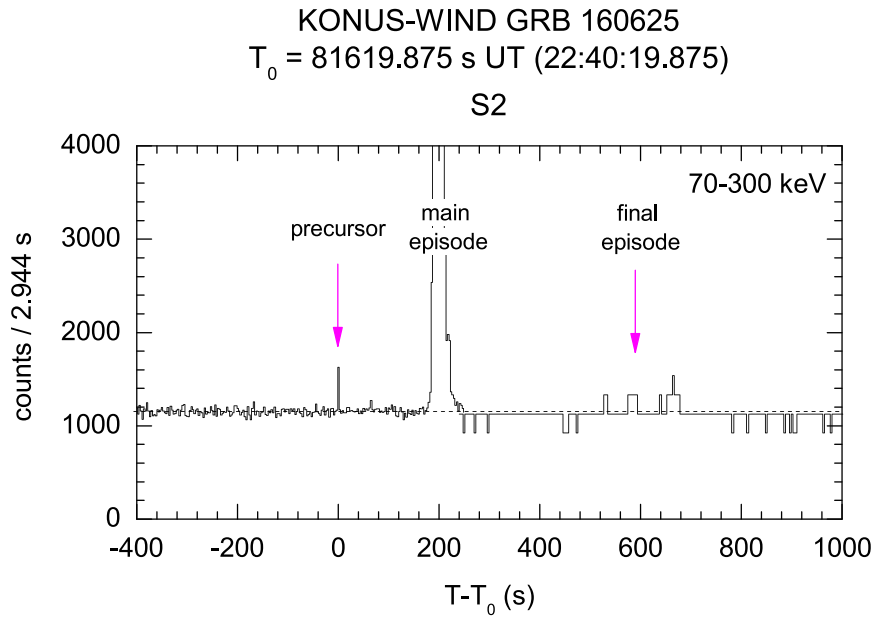

**Figure 5.** GRB 160625B light curve recorded in the KW waiting and housekeeping modes in the ~70–300 keV band (B2).

in the main episode, supporting the association of the precursor and the burst.

### 2.3.2. Time-resolved Spectral Analysis

In the triggered mode, Konus-Wind measures 64 energy spectra in 128 channels of two overlapping energy bands: 20–1170 keV (PHA1) and 244 keV–15 MeV (PHA2). The first four spectra have a fixed accumulation time of 64 ms; after that, the accumulation time varies over 0.256–8.192 s, depending on the current intensity of the burst. Five initial energy spectra covered the precursor (T0–T0+8.448 s) and 35 covered the main episode (T0+180.480–T0+237.824 s).

The spectral analysis was performed with XSPEC version 12.9.0i (Arnaud 1996) with the three models: Band GRB function (Band et al. 1993): $f(E) \sim E^{\alpha} \exp(-(2+\alpha)E/E_{peak})$ for $E < E_{peak}(\alpha - \beta)/(2+\alpha)$, and $f(E) \sim E^{\beta}$ for $E > E_{peak}(\alpha - \beta)/(2 + \alpha)$, where $\alpha$ is the power-law photon index at low energies, $E_{peak}$ is the peak energy in the $EF(E)$ spectrum, and $\beta$ is the high-energy photon index; a cutoff power-law model (CPL): $f(E) \sim E^{\alpha} \exp(-(2+\alpha)E/E_{peak})$; and a simple power law. The spectral models were normalized to the energy flux in the 20 keV–10 MeV range, a standard band for KW GRB spectral analysis. Typically, the spectral channels are rebinned to have at least 10 counts per energy bin to ensure Gaussian-distributed errors and the correctness of the $\chi^2$ statistic.

The results of the KW time-resolved spectral analysis are listed in Table 1. For the precursor, constrained spectral parameters of the CPL model are available only for the sum of the first four spectra, while for the main episode, a good count statistic is achieved for 35 individual spectra between T0+180.480 s and T0+237.824 s, the spectrum between T0+229.632 s and T0+37.824 s is well described by a simple power law. The good spectral coverage enables us to construct the temporal behavior of the model parameters ($\alpha$, $\beta$, $E_{peak}$) and to trace in detail the evolution of the spectral composition of radiation over the course of the main episode (Figure 6).

Spectrum 27 was measured at the onset of the very intense initial pulse of the main episode (Figure 6). The emission at this moment is hard; $E_{peak}$ reaches the highest value for the burst (~1.8 MeV). After the summit of the brightest pulse, $E_{peak}$ starts to decrease gradually (spectra 28–38), down to ~300 keV in spectra (39–43), and then grows to ~600 keV at the time of the second peak (spectra 44–56), staying at this level during the last peak (spectrum 57). Spectra 58–60 describe the decay of the main episode showing the typical decrease in $E_{peak}$. During the brightest part of the main episode the low-energy spectral index is approximately constant and is consistent with the synchrotron emission in the slow-cooling regime $\alpha = -2/3$ (Preece et al. 1998). The high-energy index $\beta$ shows no significant correlation with energy flux and is typical for long GRBs ($\beta \sim -2$).

Spectra 1–4 (time averaged), corresponding to the peak of the initial pulse is well fitted with a CPL model with $\alpha \sim -0.4$, and $E_{peak}$ ~70 keV ($\chi^2$/dof = 16.5/29). The fit with a single blackbody (BB) component yields kT = 17.3 (−1.6, +1.6) keV, energy flux 1.92 (−0.19, +0.19) $10^{-6}$ erg cm$^{-2}$ s$^{-1}$, with $\chi^2$/dof = 32.1/30; the fit underestimates count rate at energies below ~30 keV. Thus, despite the BB kT being consistent with the value found by Zhang et al. (2018), we argue that the BB model cannot be favored for the spectrum.

### 2.3.3. GRB 160625B in the Rest Frame

The fluence of the main episode, which accounts for ~99% of the burst fluence, is 9.36(−0.16, +0.16) $10^{-4}$ erg$^{-1}$ cm$^{-2}$ and a 256 ms peak flux, measured from T0+188.928 s, is 1.26 (−0.10, +0.10) × $10^{-4}$ erg/cm$^{2}$ s$^{-1}$ (both in the 20 keV–10 MeV energy range). The fluence of the initial pulse is ~ 1.2 × $10^{-6}$ erg cm$^{-2}$ (~$10^{-3}$ of the main episode one) and 64 ms peak flux, measured from T0+8 ms is 2.7 × $10^{-6}$ erg cm$^{-2}$ s$^{-1}$ (~$10^{-2}$ of the main episode one).

Assuming the burst redshift $z = 1.406$ (Xu et al. 2016) and a standard cosmology model with $H_0 = 67.3$ km s Mpc$^{-1}$, $\Omega_M = 0.315$, and $\Omega_\Lambda = 0.685$ (Planck Collaboration 2014),

Table 1
Konus-Wind Time-resolved Spectral Fits

| Spectrum | Accumulation Interval (s from T0) | $\alpha$ | $\beta$ | $E_{\text{peak}}$ (keV) | Flux $10^{-6}$ erg cm$^{-2}$ s$^{-1}$ | $\chi^2$/dof |
|---|---|---|---|---|---|---|
| 1-4 | 0.000–0.256 | $-0.38^{+0.68}_{-0.57}$ | ... | $67^{+8}_{-8}$ | $2.1^{+0.2}_{-0.2}$ | 16.5/29 |
| 27 | 180.480–187.904 | $-0.93^{+0.08}_{-0.07}$ | $-2.01^{+0.20}_{-0.40}$ | $1828^{+564}_{-429}$ | $7.8^{+0.7}_{-0.7}$ | 99.8/97 |
| 28 | 187.904–188.160 | $-0.71^{+0.15}_{-0.12}$ | $-1.96^{+0.15}_{-0.28}$ | $1030^{+425}_{-281}$ | $94.9^{+8.1}_{-8.1}$ | 97.5/67 |
| 29 | 188.160–188.416 | $-0.59^{+0.18}_{-0.12}$ | $-2.07^{+0.15}_{-0.20}$ | $826^{+229}_{-221}$ | $115.7^{+8.4}_{-8.4}$ | 92.6/67 |
| 30 | 188.416–188.672 | $-0.69^{+0.12}_{-0.09}$ | $-2.25^{+0.18}_{-0.25}$ | $940^{+220}_{-202}$ | $128.2^{+8.6}_{-8.6}$ | 82.4/69 |
| 31 | 188.672–188.928 | $-0.69^{+0.14}_{-0.11}$ | $-2.26^{+0.18}_{-0.33}$ | $951^{+305}_{-220}$ | $132.4^{+8.8}_{-8.8}$ | 102.7/70 |
| 32 | 188.928–189.184 | $-0.77^{+0.10}_{-0.08}$ | $-2.22^{+0.17}_{-0.25}$ | $937^{+195}_{-179}$ | $126.3^{+9.0}_{-9.1}$ | 62.0/67 |
| 33 | 189.184–189.440 | $-0.61^{+0.14}_{-0.12}$ | $-2.07^{+0.11}_{-0.15}$ | $644^{+159}_{-116}$ | $113.5^{+8.0}_{-8.0}$ | 67.8/67 |
| 34 | 189.44–189.696 | $-0.57^{+0.16}_{-0.13}$ | $-2.02^{+0.11}_{-0.14}$ | $529^{+133}_{-110}$ | $95.6^{+7.2}_{-7.2}$ | 86.6/67 |
| 35 | 189.69–189.952 | $-0.58^{+0.17}_{-0.17}$ | $-2.09^{+0.12}_{-0.21}$ | $515^{+179}_{-102}$ | $71.4^{+6.1}_{-6.1}$ | 87.7/62 |
| 36 | 189.952–190.208 | $-0.55^{+0.20}_{-0.18}$ | $-2.00^{+0.10}_{-0.14}$ | $422^{+120}_{-81}$ | $60.1^{+5.7}_{-5.7}$ | 75.7/63 |
| 37 | 190.208–190.464 | $-0.82^{+0.20}_{-0.14}$ | $-2.43^{+0.28}_{-0.67}$ | $577^{+193}_{-163}$ | $42.5^{+4.7}_{-4.7}$ | 67.1/58 |
| 38 | 190.464–190.720 | $-0.76^{+0.19}_{-0.16}$ | $-2.13^{+0.16}_{-0.27}$ | $453^{+157}_{-101}$ | $37.6^{+4.7}_{-4.6}$ | 85.0/56 |
| 39 | 190.72–191.232 | $-0.76^{+0.17}_{-0.13}$ | $-2.19^{+0.15}_{-0.22}$ | $401^{+99}_{-84}$ | $29.7^{+3.0}_{-2.9}$ | 57.9/63 |
| 40 | 191.232–191.744 | $-0.51^{+0.29}_{-0.19}$ | $-2.27^{+0.17}_{-0.26}$ | $314^{+76}_{-72}$ | $23.0^{+2.6}_{-2.5}$ | 96.0/61 |
| 41 | 191.744–192.256 | $-0.74^{+0.18}_{-0.18}$ | $-2.44^{+0.20}_{-0.56}$ | $342^{+108}_{-62}$ | $19.5^{+2.2}_{-2.2}$ | 83.0/58 |
| 42 | 192.256–192.768 | $-0.72^{+0.18}_{-0.14}$ | $-2.37^{+0.19}_{-0.31}$ | $367^{+78}_{-68}$ | $22.7^{+2.5}_{-2.4}$ | 59.0/62 |
| 43 | 192.768–193.280 | $-0.74^{+0.11}_{-0.10}$ | $-2.63^{+0.22}_{-0.46}$ | $425^{+65}_{-52}$ | $28.8^{+2.6}_{-2.5}$ | 74.1/60 |
| 44 | 193.280–193.536 | $-0.76^{+0.14}_{-0.12}$ | $-2.74^{+0.36}_{-0.98}$ | $518^{+99}_{-93}$ | $34.7^{+4.0}_{-3.8}$ | 59.4/55 |
| 45 | 193.536–193.792 | $-0.70^{+0.11}_{-0.10}$ | $-2.85^{+0.33}_{-0.95}$ | $568^{+82}_{-66}$ | $41.1^{+4.3}_{-4.3}$ | 60.2/57 |
| 46 | 193.792–194.048 | $-0.64^{+0.13}_{-0.11}$ | $-3.14^{+0.50}_{+3.14}$ | $544^{+77}_{-73}$ | $39.5^{+4.2}_{-4.0}$ | 59.4/57 |
| 47 | 194.048–194.304 | $-0.67^{+0.12}_{-0.11}$ | $-2.62^{+0.28}_{-0.56}$ | $616^{+107}_{-94}$ | $49.7^{+5.1}_{-5.1}$ | 69.4/59 |
| 48 | 194.304–194.560 | $-0.76^{+0.14}_{-0.12}$ | $-2.78^{+0.38}_{-1.47}$ | $597^{+132}_{-115}$ | $45.1^{+4.6}_{-4.6}$ | 54.1/58 |
| 49 | 194.560–194.816 | $-0.52^{+0.17}_{-0.15}$ | $-2.54^{+0.21}_{-0.39}$ | $448^{+94}_{-72}$ | $45.9^{+4.3}_{-4.3}$ | 80.6/58 |
| 50 | 194.816–195.072 | $-0.75^{+0.14}_{-0.12}$ | $-2.65^{+0.29}_{-0.71}$ | $529^{+112}_{-90}$ | $45.4^{+4.6}_{-4.5}$ | 72.9/58 |
| 51 | 195.072–195.328 | $-0.61^{+0.16}_{-0.14}$ | $-2.43^{+0.20}_{-0.34}$ | $405^{+77}_{-63}$ | $39.3^{+4.3}_{-4.2}$ | 43.0/56 |
| 52 | 195.328–195.584 | $-0.72^{+0.12}_{-0.11}$ | $-2.52^{+0.23}_{-0.41}$ | $518^{+89}_{-80}$ | $47.8^{+4.9}_{-4.8}$ | 41.7/56 |
| 53 | 195.584–195.840 | $-0.66^{+0.14}_{-0.12}$ | $-2.63^{+0.29}_{-0.63}$ | $499^{+89}_{-78}$ | $39.4^{+4.5}_{-4.3}$ | 49.2/55 |
| 54 | 195.840–196.096 | $-0.70^{+0.19}_{-0.15}$ | $-2.44^{+0.24}_{-0.47}$ | $415^{+102}_{-82}$ | $33.5^{+4.2}_{-4.0}$ | 44.0/55 |
| 55 | 196.096–196.352 | $-0.79^{+0.13}_{-0.11}$ | $-3.41^{+0.77}_{+3.41}$ | $567^{+89}_{-92}$ | $32.3^{+3.9}_{-3.5}$ | 49.6/55 |
| 56 | 196.352–196.864 | $-0.72^{+0.13}_{-0.12}$ | $-2.80^{+0.29}_{-0.67}$ | $366^{+53}_{-45}$ | $20.9^{+2.0}_{-1.9}$ | 51.4/59 |
| 57 | 196.864–205.056 | $-0.79^{+0.02}_{-0.02}$ | $-2.74^{+0.10}_{-0.12}$ | $612^{+23}_{-22}$ | $35.4^{+0.8}_{-0.8}$ | 187.2/97 |
| 58 | 205.056–213.248 | $-0.94^{+0.06}_{-0.06}$ | $-2.57^{+0.17}_{-0.27}$ | $338^{+30}_{-27}$ | $7.0^{+0.5}_{-0.5}$ | 100.5/97 |
| 59 | 213.248–221.440 | $-1.07^{+0.29}_{-0.20}$ | $-2.29^{+0.26}_{-0.77}$ | $219^{+74}_{-60}$ | $1.4^{+0.3}_{-0.3}$ | 116.8/97 |
| 60 | 221.440–229.632 | $-1.12^{+0.56}_{-0.38}$ | ... | $321^{+520}_{-132}$ | $0.3^{+0.2}_{-0.1}$ | 50.1/61 |
| 61 | 229.632–237.824 | $-1.95^{+0.39}_{-0.56}$ | ... | ... | $0.2^{+0.3}_{-0.1}$ | 59.3/62 |

**Note.** Time-integrated spectrum of the initial pulses (1–4) and spectrum 60 is fitted with the CPL model, spectra 27–59 are fitted with the Band function, and spectrum 61 is fitted by a simple power law.

we estimate the rest-frame parameters of the episodes, see Table 2 for the observed and rest-frame energetics.

The burst isotropic energy release $E_{\text{iso}}$ of $\sim 5.0 \times 10^{54}$ erg is the second highest one measured by Konus-Wind for more than 25 yr of its GRB observations (the highest one $\sim 5.8 \times 10^{54}$ erg was measured for GRB 090323, z = 3.6); the peak luminosity $L_{\text{iso}}$ is $\sim 1.6 \cdot \times 0^{54}$ erg s$^{-1}$, and the rest-frame peak energy of the time-integrated ($E_{p,i,z}$) and peak spectrum ($E_{p,p,z}$) are $\sim 1.3$ and $\sim 2.3$ MeV, respectively.

With these values, GRB 160625B is a *typical* long GRB (Figure 7), and it lies within 90% of the prediction bands for both the Amati and Yonetoku relations built for the sample of 138 long KW GRBs with known redshifts (Tsvetkova et al. 2017). The initial pulse is also consistent with the relations, which may indicate the single emission mechanism in the precursor and main episode.

### 3. Quasiperiodic Optical Variability of the Intrinsic Optical Radiation GRB 160625B

A detailed consideration of the GRB 160625B optical light curve (Figure 8) led to the assumption of the existence of a quasiperiodic shine fluctuation with a period of the order of 20 s. In this section, we consider statistical criteria that support this assumption.

First of all we need to eliminate the general trend of light curve. The decrease in the prompt GRB emission in the optical range can be described by a power law with index k, which undergoes an abrupt change at the time $T_k = T_{\text{LAT}} + 85$ s. The coefficients are determined by the least squares method. Thus, in the interval of 30–85 s, the value of k = 0.89, $\sigma_k = 0.06$, and then changes to k = 3.33, $\sigma_k = 0.07$.

The section of the light curve, starting from Tk, is also well described by the exponential decline (linear dependence in stellar magnitudes—$m = B^*t + A$)–$A = 9.27$, $\sigma_A = 0.07$; $B = 0.0072$ $\sigma_B = 0.0004$.

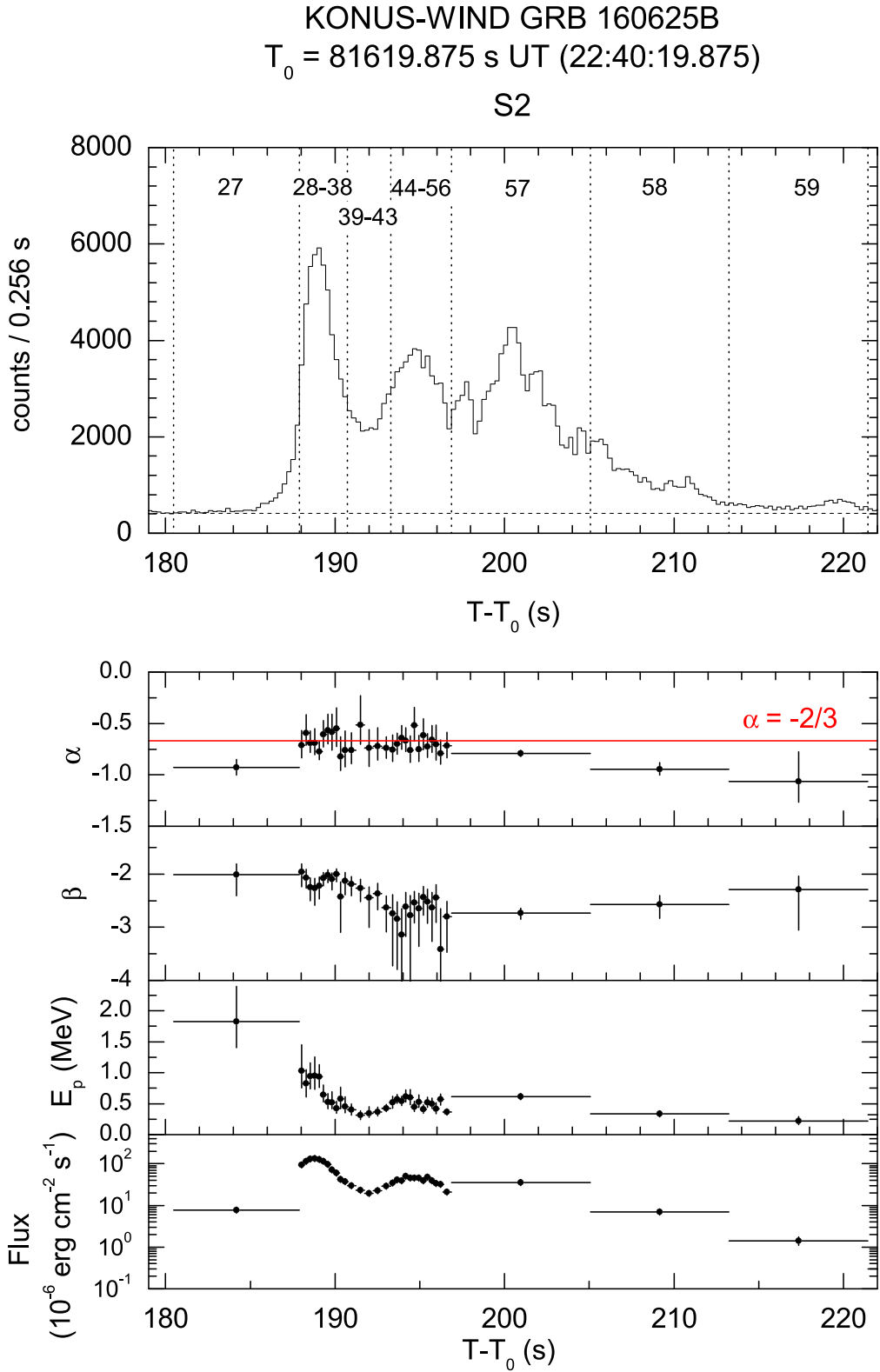

**Figure 6.** Spectral evolution of the gamma-ray emission during the prompt phase of the burst main episode. The Konus-Wind light curve in the combined B1+B2+B3 energy band (17–1170 keV) is shown with 256 ms resolution, along with the temporal behavior of the Band spectral model parameters $E_{peak}$, $\alpha$, and $\beta$ obtained from the time-resolved fits (see Table 1).

To understand whether a deterministic component is present in the time series obtained after subtracting the trend, we use an Abbe test (see Ajvazyan et al. 1983). The Abbe criterion consists of testing the hypothesis of the independence of the members of a numerical series (null hypothesis), assuming that all of them are a sample from a normal general set. The essence of the Abbe method (or the method of successive differences) is to compare variances calculated in different ways. If we denote by $\bar{x}$ the arithmetic mean of all the terms of the series (after subtracting the trend), then the usual expression of the variance is

$$\sigma^2(N) = \frac{1}{n-1}\sum_{n=1}^{N}(x_n - \bar{x})^2 \quad (1)$$

Table 2
Observed and Rest-frame Energetics of the Initial Pulse and the Main Episode

| Episode | Fluence $10^{-6}$ erg cm$^{-2}$ | $E_{\rm iso}$ $10^{52}$ erg | Peak Flux $10^{-6}$ erg cm$^{-2}$ s$^{-1}$ | $L_{\rm iso}$ $10^{52}$ erg s$^{-1}$ | $E_{p,i,z}$ keV | $E_{p,p,z}$ keV |
|---|---|---|---|---|---|---|
| 1 | $1.25^{+0.14}_{-0.13}$ | $0.67^{+0.07}_{-0.07}$ | $2.7^{+0.5}_{-0.5}$ | $3.5^{+0.6}_{-0.6}$ | $161^{+19}_{-19}$ | $161^{+19}_{-19}$ |
| 2 | $936^{+16}_{-16}$ | $500^{+9}_{-9}$ | $126.0^{+9.7}_{-9.7}$ | $162^{+12}_{-12}$ | $1332^{+47}_{-45}$ | $2256^{+469}_{-431}$ |

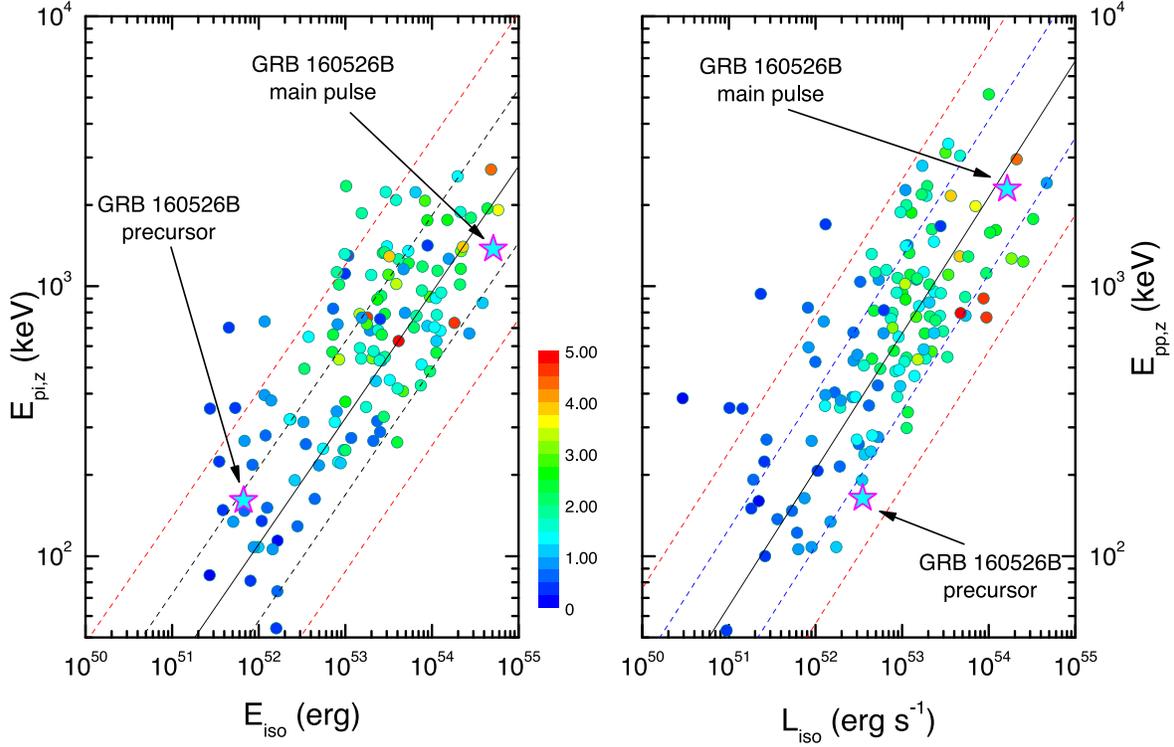

**Figure 7.** GRB 160625B in the $E_{p,i;z}$–$E_{\rm iso}$ and $E_{p,p;z}$–$L_{\rm iso}$ planes. The rest-frame parameters of the precursor and the main episode are indicated by stars. The rest-frame parameters of 138 long KW GRBs with known redshifts detected by KW in the triggered mode (Tsvetkova et al. 2017) are shown with circles and their best-fit Amati and Yonetoku relations are plotted with solid lines; the dashed lines denote their 68% and 90% prediction bounds. The color of each data point represents the burst's redshift.

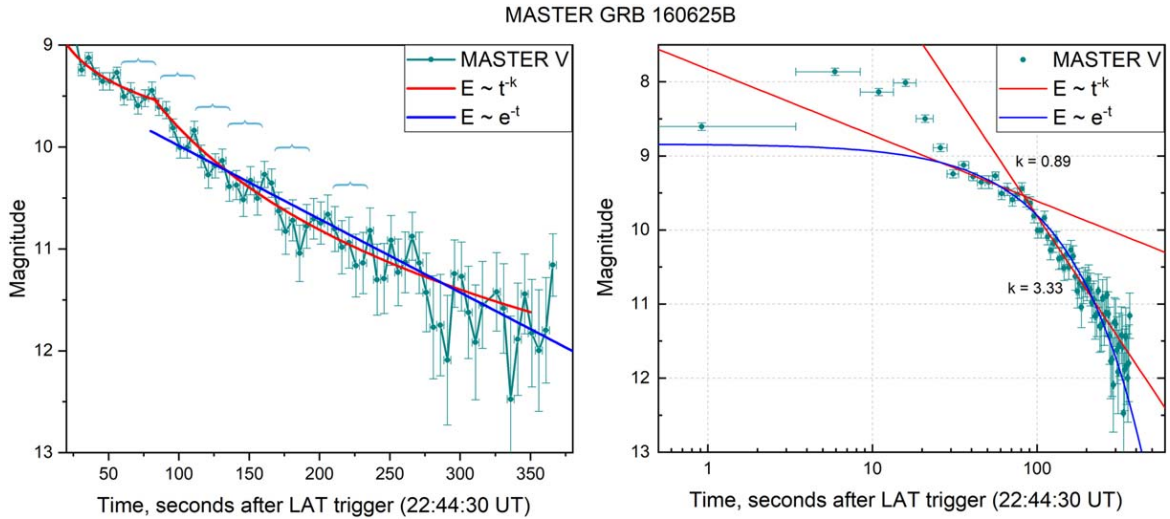

**Figure 8.** Fitting the light curve under the assumption of exponential decay of the flux—blue curve, as well as under the assumption of a flux drop $\sim t^{-k}$, where at time $T_k = 85$ s after the LAT trigger there was a sharp change in the parameter $k$—red straight lines.

Here, and further on, $N$ is the number of terms of the series.

The variance can be estimated in a different way

$$Q^2(N) = \frac{1}{2(n-1)}\sum_{n=1}^{N-1}(x_{n+1} - x_n)^2 \quad (2)$$

Here, the squares of the increments of the flux measured at adjacent time moments are summed.

Note that the value of $Q^2$ is weakly affected by the deterministic component, since we take the difference in the neighboring terms of the series.

If the series contains a systematically changing component, then for each moment of time there is its own mean value. By definition, Equation (1) includes the average $\bar{x}$ common to the entire series. Therefore, the variance calculated by this formula will be greater than that estimated by formula (2), which is weakly dependent on the presence of a deterministic component.

The criterion for the presence of a deterministic component is the ratio $\gamma = Q^2(N)/\sigma^2(N)$.

In other words, the question of the statistical independence of the members of the studied series, i.e., the question of the absence or presence of systematic changes is solved by the inequality

$$\gamma < \gamma_c(N). \quad (3)$$

Here, $\gamma_c$ is a critical value, whose magnitude is tabulated for different values of N and for different significance levels. For $N > 60$, there are formulas for estimating the value of $\gamma_c$, for example (see Bolshev & Smirnov 1968),

$$\gamma_c(N) = 1 - \frac{u_\alpha}{\sqrt{N + 0.5(1 + u_\alpha^2)}}. \quad (4)$$

A quantity $u_\alpha$ is the quantile of the normalized normal distribution, and $\alpha$ is the statistical significance level.

If inequality (3) is satisfied, then the null hypothesis is rejected and the elements of our sample cannot be considered statistically independent, i.e., we should accept an alternative hypothesis about the presence of nonrandom changes in detrended radiation flux. Simple calculations give the value of the observed statistic at $N = 67: \gamma \approx 0.70$ (two-component curve) and at $N = 55: \gamma \approx 0.67$ (strict line). The critical value of $\gamma_c$ (67) at the significance level of $\alpha = 0.01$ is $\gamma_c$ (67) $\approx 0.72$ and inequality (3) is satisfied. For linear regression, $\gamma_c$ (55) $\approx$ 0.695 at the significance level of $\alpha = 0.01$ and inequality also is satisfied. By modifying the Abbe test using Poisson noise, we obtain similar results (see Figure 12 in the Appendix). In this case, the Abbe inequality is satisfied with a significance level of $\alpha = 0.023$ (two-component curve). Therefore, we must reject the null hypothesis and accept that the light curve after subtracting the trend contains the deterministic (possibly quasiperiodic) component.

To estimate the characteristic time period of the assumed (quasi)periodic deterministic component, we use the Lafler–Kinman method (1965). This is a continuation of the Abbe method and allows studying time series for periodicity. Without going into the details, which can be found in the original works, or in the book by Terebizh (1982) that describes a wide range of techniques for studying time series, we only note that this method examines the phase diagram, and not the original series $M(t_k)$. Each term of the original time series is assigned a phase:

$$X_k(\nu) = fr(\nu(t_k - t_0)). \quad (5)$$

Here, $fr(\nu(t_k - t_0))$ is a fractional part of the number, $\nu$ is a trial frequency (quasiperiodic) of the deterministic component, and $t_0$ is an arbitrary initial time.

All the terms of the new series are ordered in ascending order $X_k$, and then the Abbe statistics are calculated for a certain set of trial frequencies.

Obviously, if there is a periodic component, then at its frequency, the Abbe parameter will be minimal. Thus, we found that the periodogram shows several peaks. The strongest of them correspond to periods of 25 and 50 s.

Another interesting feature is the visually noticeable increase in the time between brightness fluctuations (Figure 8); a possible explanation for this phenomenon and the physical reason for all quasiperiodicity will be discussed in Section 5.

## 4. Comparison of Optical and Gamma-Ray Spectrum Data

Below are images of the spectrum according to the data of the Konus-Wind observatory. Here, along with the curve described by the Band function model (Band et al. 1993), the data from optical observations made by the MASTER telescopes are plotted. It can be seen that the high-frequency part of the spectrum (kiloelectronvolt–megaelectronvolt) fits well with the observational data, but the optical flux is almost four orders of magnitude higher than what is prescribed for a standard curve extrapolated to the low-frequency region. This excess disappears after 30 s after reaching the maximum of the light curve. That is, we see an optical flash whose duration approximately coincides with the duration of the fast emission phase (prompt emission). At first glance, this coincidence of durations indicates that gamma and optical radiation are born in the same place. However, as emphasized in Zhang et al. (2018), the optical light curve somewhat lags behind changes in the gamma range for about 3 s.

This lag can be interpreted as a confirmation of optical radiation being generated far away from the gamma's birth site, at distances of the order of $10^{17}$ cm. In this interpretation, the optical flash is a manifestation of an external reverse shock, i.e., the sources of optical and gamma radiation are spaced apart.

It is known that the afterglow of the reverse shock in the IR and radio range can be observed quite long after the main event —the gamma-ray flash (Sari & Piran 1999; Kobayashi 2000). By studying the behavior of the light curve in the radio range, it is possible to estimate the main physical parameters that characterize the reverse shock wave and determine the radiation flux. This kind of analysis was performed in Lü et al. (2017). Based on the estimates of the physical parameters of the shock wave obtained from late radio observations, the authors synthesized the optical light curve and extrapolated it back in time. As a result, the gap between the observed flow in optics and the flow that is dictated by the extension of the standard model of the spectrum to the optical frequency range is reduced, but not completely eliminated.

**Table 3**
Correspondence of KW Spectra and MASTER Exposures, KWsp—KW Spectrum Number

| KWsp | KWspTi | KWspTf | k | N | MR_T | MR_Exp | MR_Vmag | MR_VmagErr | MR_TKW | MR_Ti | MR_Tf |
|---|---|---|---|---|---|---|---|---|---|---|---|
| 28–40 | 187.904 | 191.744 | 1.00 | 1 | 0.92 | 5 | 8.6 | 0.05 | 189.23885 | 186.739 | 191.739 |
| 41–56 | 191.744 | 196.864 | 1.00 | 2 | 5.92 | 5 | 7.86 | 0.05 | 194.23885 | 191.739 | 196.739 |
| 57 | 196.864 | 205.056 | 1.00 | 3 | 10.92 | 5 | 8.14 | 0.05 | 199.23885 | 196.739 | 201.739 |
| 58 | 205.056 | 213.248 | 0.34 | 4 | 15.92 | 5 | 8.01 | 0.05 | 204.23885 | 201.739 | 206.739 |
| 58 | 205.056 | 213.248 | 1.00 | 5 | 20.92 | 5 | 8.5 | 0.05 | 209.23885 | 206.739 | 211.739 |
| 59 | 213.248 | 221.440 | 0.7 | 6 | 25.92 | 5 | 8.89 | 0.05 | 214.23885 | 211.739 | 216.739 |
| 60 | 221.440 | 229.632 | 0.06 | 7 | 30.92 | 5 | 9.24 | 0.05 | 219.23885 | 216.739 | 221.739 |

**Note.** Record type 28–40 means the sum of spectra from the 28th to 40th inclusive. KWspTi, KWspTf—beginning and end of the spectrum accumulation interval relative to T0KW, k—MASTER exposure fraction falling into the corresponding KW spectrum, N—MASTER exposure ordinal number, MR_T—MASTER exposure middle time relative to T0$_{LAT}$, MR_Exp, MR_Vmag, MR_VmagErr—exposure time, stars magnitude, and its error, MR_TKW is the time of the middle of the MASTER exposure relative to T0KW = MR_T + 188.31885 s, MR_Ti, MR_Tf are the start and end times of the MASTER exposure relative to T0KW.

Long bright GRBs like GRB 160625B are rare, but not unique. An example of this is GRB 080319B (Racusin et al. 2008). It also has a significant gap between the optical flux and the flux obtained by extrapolating the standard spectrum to the low-frequency range. Investigating the nature of GRB 080319B, Kumar & Narayan (2009) (see also Lazar et al. 2009) proposed a relativistic turbulence model to explain the fine structure of the light curve observed in the gamma range. It is assumed that the matter of the dropped shell is divided into two phases: the turbulent cells and everything in between. In this case, gamma radiation is generated in both phases (reverse Compton), and synchrotron radiation of electrons in the inter-cell space is responsible for the radiation flux in the optics.

This explains a number of observed properties of GRBs, and what is important in the context of this work, the huge excess of the flux in the optical range over what is dictated by the Band function. However, the natural consequence of this theory is a diminishing of the polarization over time, since the assumed turbulence entangles the magnetic field, leading to depolarization. In fact, the MASTER polarization observations (see Troja et al. 2017) show an increase in polarization. Thus, in the case of GBR 160625B, the relativistic turbulence model does not work.

In summary, we can conclude that the optical flash is quite likely to be a manifestation of an external reverse shock wave; however, the issue requires further study.

The spectra were plotted using the HEASoft software package (https://heasarc.gsfc.nasa.gov/docs/software/heasoft/). The time intervals were selected for which the MASTER and Konus-Wind observations intersected (Table 3), after which the data on the joint plot were approximated by the GRB model (https://heasarc.gsfc.nasa.gov/xanadu/xspec/manual/node181.html) using the XSPEC utility.

## 5. Modeling Hypotheses

Looking at the optical light curve starting from 20 s relative to the main pulse and up to 200–250 s, one can suspect quasiperiodic brightness oscillations with an amplitude significantly exceeding the random brightness measurement error. Recall that a 5 s exposure on ultra-wide-field cameras is practically without delay. Such variability at times of the order of ($\tau \sim 10$–20 s) is caused by internal physical processes in the operation of the central engine. However, it would be tempting to associate this phenomenon with the interaction of a relativistic jet with quasiperiodic inhomogeneous layers in the radial distribution of the progenitor's stellar wind . This type of inhomogeneity can be caused by the presence of a close second component in the collapsing star's binary system. Calculations carried out earlier (Lipunova et al. 2009) show that in systems with an orbital period of less than 1–2 hr, tidal forces lead to a critical increase in the spin moment of the collapsar. As a result, it is under these conditions that the formation of a massive accretion disk (Woosley 1993) or a spinar (Lipunov & Gorbovskoy 2007) prolongs the magneto-rotational collapse, which leads to the phenomenon of a GRB.

First, we will discuss the possible reasons for the quasiperiodic optical emission, which is an intrinsic emission at the stage of the central engine operation and must reflect some quasiperiodic oscillations of the central engine. This circumstance is supported by the fact that we detect oscillations after 30 s (in our o frame of reference) from the trigger, that is, at the same time when the optical and gamma spectra become one (Figure 9). Recently, Suvorov & Kokkotas (2020) and Suvorov & Kokkotas (2021) discussed quasiperiodic pulsations of X-rays from some short GRBs in the magnetar model. Actually, the only property of the magnetars used here is the one inherent to all radio transmitters; they lose energy approximately according to the magnetic dipole law (Pacini 1967) and their total power changes over time $L \sim t^{-2}$. In connection with GRBs, this circumstance was expressed by Lipunova & Lipunov (1998). The merit of the works of Suvorov & Kokkotas (2020), Suvorov & Kokkotas (2021) is the attraction of the idea that the observed quasiperiodic oscillations are associated with the free precession of a neutron star with an anomalously strong magnetic field. In fact, the magnetar model is only a special case of a more general model of the magneto-rotational collapse—the spinar paradigm (Lipunova & Lipunov 1998; Lipunov & Gorbovskoy 2007, 2008; Lipunova et al. 2009). In the framework of the spinar paradigm, it is assumed that the initial rotational moment of the body is so great that centrifugal forces has a significant effect on the collapse process. In particular, this model successfully explains not only the plateau phenomenon, but also a sharp (by several orders of magnitude) cliff at the end of the plateau (Lipunov & Gorbovskoy 2007). In fact, an approximate nonstationary model of gravitational-rotational collapse (Lipunov & Gorbovskoy 2008) includes all relativistic effects plus the contribution of the nuclear forces of the neutron liquid. The collapse character depends on three main parameters: the core mass M of the collapsing star, the generalized Kerr parameter $a_0 = I\omega_0 c/GM^2$, and the ratio of magnetic and gravitational energy $\alpha_m = U_m/U_{gr} \ll 1$, which remains constant in the approximation of the conservation of the magnetic flux). A black hole or a neutron star can also be the final product of the collapse (Lipunova et al. 2009).

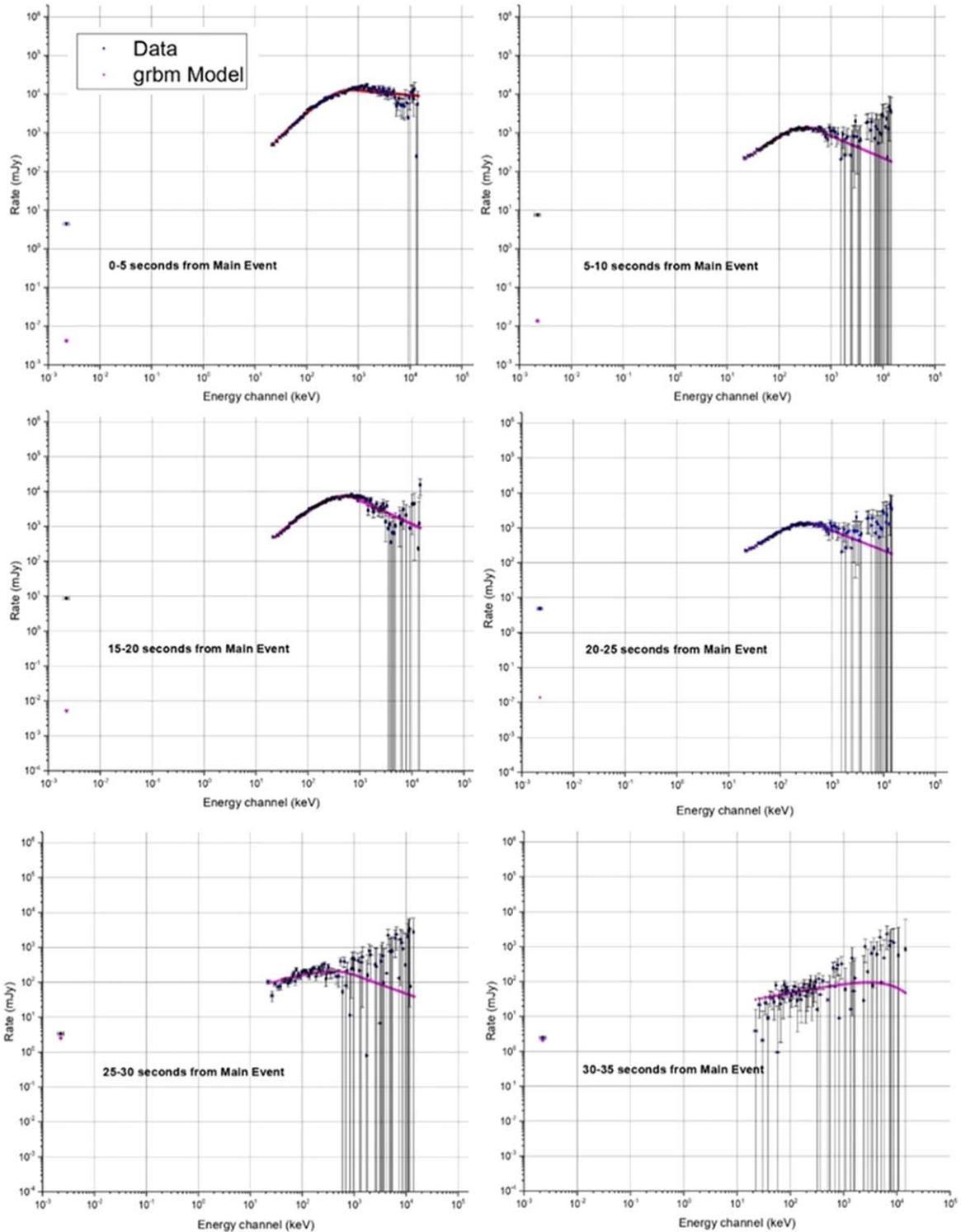

**Figure 9.** Spectra built using the HEASoft package in the intervals of the intersection of the MASTER and Konus-Wind observations (Table 1), approximated by the phabs*zphabs*grbm model using the XSPEC utility. The formation of a single spectrum occurs at a time interval of 25–30 s after the LAT trigger at 22:43:24 UT. Spectral parameters $\alpha$, $\beta$-photon indices, $E$-characteristic energy.

In the process of the formation of both types of objects, an intermediate object is formed (a spinar), which can experience not only free, but also forced precession. In addition, if a neutron star can only slow down, decreasing the overall luminosity, the spinar can accelerate as it evolves, and this will even be accompanied at certain stages by an increase in luminosity. It is precisely by the spinar precession that we propose to explain the suspected oscillations of the intrinsic optical emission of the GRB 160625B gamma burst. Remarkably, the magneto-rotational collapse model naturally explains the existence of the precursor, with which we begin.

In the spinar model, the precursor of the GRBs occurs during the first abrupt stop of the collapse due to the increase in centrifugal force (Figure 10).

At this moment, half of the gravitational energy accumulated by the free-falling progenitor core is converted into heat. The second half of the work of gravitational forces turns into the energy of the spinar rotation. In other words, the centrifugal forces stop the collapse and a spinar is formed—which continues to slowly contract, losing its extreme moment and constantly increasing its radiation power (see Figure 10). This is the formation of the precursor! Its maximum power $E_{\text{precursor}} \approx GM^2/2R_{\text{sp}}$ is determined by the radius of the formed spinar (Lipunov & Gorbovskoy 2007) $R_{\text{sp}} \approx a_0^2 R_g/2$, where $R_g$ is the Schwarzschild radius:

$$E_{\text{precursor}} \approx GM^2/2R_{\text{sp}} \approx (1/2a_0^2)Mc^2. \quad (6)$$

If the core torque is not too high, $a_0 < 5$–$10 m^{-1}$ ($m = M/M_\odot$), then this energy directed mainly along the spinar rotation axis will pierce the progenitor shell and we will see the precursor (Lipunov & Gorbovskoy 2007).

However, immediately after the formation of the spinar, the energy released by the spinar will already be determined by the magnitude of the magnetic energy $L_{\text{sp}} = U_m\omega \approx \alpha_m U_{gr}\omega$. Accordingly, the energy released by the spinar in one spin period and the corresponding pressure impulse on the shell will be reduced by a factor of $\alpha_m \ll 1$; hence, the jet breakdown will happen. But if the torque is lost, the luminosity of the spinar will grow $L_{\text{sp}} = U_m\omega = \epsilon_m U_{gr}\omega$ and finally break through the shell. This will be the time close to the beginning of the GRB $t_{\text{GRB}}$. However, the time elapsed from the precursor to the start of the GRB will be determined by the rate of loss of the spinar torque at the moment of its formation: $\Delta t \approx I\omega/U_m$. We can express the magnetic energy through the magnetic flux $\Phi = \pi BR^2 = \Phi_{28} 10^{28}$ Gs cm$^2$, thus normalizing the flux to the characteristic of magnetars does not change during the collapse, according to our assumption. Then we get the precursor time:

$$\Delta t_1 \approx -800 s \Phi_{28}^{-2} m_{10}^3 a_0^3. \quad (7)$$

Kerr black hole released energy can reach 42% (Thorne 1974), and we get another important ratio of precursor fluence to GRB fluence $E_{\text{pre}}/E_{\text{GRB}} \approx a_0^{-2}$. Usually there is a ratio of $E_{\text{pre}}/E_{\text{GRB}} \approx 1\%$–$10\%$ (Troja et al. 2007), corresponding to the generic Kerr parameter $a_0 \sim 3$–$10$. For the masses of nuclei close to $M \sim$ MOV, a larger Kerr parameter is required. However, in this case, it is necessary to include the contribution of nuclear forces (Lipunova et al. 2009).

After the formation of the spinar, the direction of the magnetic flux of course does not have to coincide with its axis of rotation. We recall that we are considering the case of the conservation of the magnetic flux without an accompanying generation of the type of dynamo mechanism. Therefore, a spinar can participate in a free precession, especially when it turns into a magnetar. However, a spinar is not a neutron star, whose surface is a hard boundary, beyond which there is a vacuum, or a highly discharged magnetosphere with a Julian–Goldreich density (Lipunov 1992). Spinar is an idealization of a superdense rotating body surrounded by a gas-dynamic plasma. Its evolution can be described by the equation

$$dI\omega/dt = K_\parallel + K_\perp. \quad (8)$$

On the right is the moment of forces, the parallel component of which leads to a change in the spinar's rotational moment in magnitude, and the perpendicular component leads to forced precession. It is clear that the spinar is not a rigid body, but the experience of studying the precession of accretion disks under the action of the magnetic moments of the seat shows that even thin disks successfully precess, though in a differential way (Lipunov et al. 1980). The maximum value of both moments of forces in Equation (3) is determined as $|\mathbf{K}| \approx Um = \epsilon_m U_{gr}$. The forced precession frequency turns out to be of the order of $\Omega = |\mathbf{K}_\perp \mathbf{I}| \approx \epsilon_m \Omega$. As the spinar radius approaches the event horizon $R \to R_g/2$ spinar frequency tends to the $\Omega \to c/R_g$ as $R \to R_g$. So far as $U_m \to \epsilon_m Mc^2$ we get an estimate of the precession period

$$T = 2\pi/\Omega = (2\pi/\omega)\epsilon m^{-1} \approx 5000\,sec\,\Phi_{28}^{-2} m_{10}^3 r^{3/2}. \quad (9)$$

Here, $r = R/R_g$.

For the precursor GRB160725B in its own frame of reference $\Delta t_{\text{pre}} \approx -70$ s (we took $z = 1.406$ (Xu et al. 2016), and the characteristic time of variations is $T \approx 10$ s. From Equations (2) and (4) we obtain the estimate $a \approx 10$ and $\Phi_{27}^{-2} m^3 \approx 140$. Accordingly, the magnetic flux turns out to be quite reasonable: $\Phi_{27}^2 \approx 140\, m^{3/2}$.

After the main pulse (G2), the luminosity of the spinar begins to decrease. This is a sure sign that the role of nuclear forces is becoming important. The role of nuclear forces can become important only if the mass of the collapsing nucleus does not greatly exceed the Oppenheimer–Volkov limit. Let us recall that the Oppenheimer–Volkov limit essentially depends on the contribution of centrifugal forces to the equilibrium of the neutron star. So, if the fraction of the rotational energy in the virial theorem is 10%–20%, the Oppenheimer–Volkov limit can increase by 2–3 times, depending on the equation of state of the neutron star (Lipunov 1992). With a mass of 4–5 $M_\odot$, the spinar will be supported by nuclear forces until it freezes. This will happen during the after

$$\Delta t_3 \approx I\omega U_m \approx 6I\omega R/\Phi 2 \approx 10 sm^2 \Phi_{27}^{-2} r^2.$$

So far as $r \to 1/2$ relativistic effects should be taken into account immediately.

Next, we applied a dynamic model of gravimagnetic collapse, which allows us to describe the evolution of a collapsing spinar from the moment of the loss of stability to the collapse into a black hole (Lipunov & Gorbovskoy 2008; Lipunova et al. 2009). In this formulation, a simple nonstationary three-parameter model of collapse is obtained with the decisive role of the rotation and the magnetic field. The input parameters of the theory are the mass, angular momentum, and magnetic field of the collapsar.

The approximate model includes: centrifugal force, relativistic effects of the Kerr metric, pressure of nuclear matter, dissipation of angular momentum under the influence of a spinar magnetic field, decrease in the dipole magnetic moment as a result of compression and the effects of general relativity (a black hole has no hair), neutrino cooling, and time dilation due to gravitational redshift.

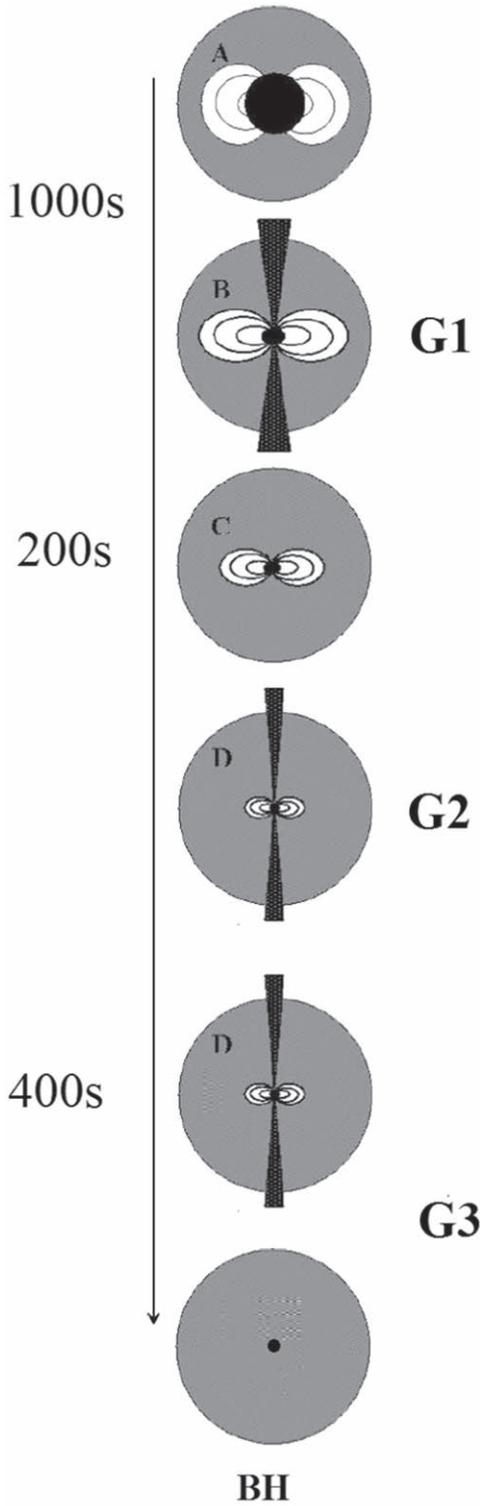

**Figure 10.** Schematic of the three-stage magneto-rotational collapse in the spinar paradigm.

In Figure 11, the results of calculating our approximate model are shown, the parameters of which are given in Table 1.

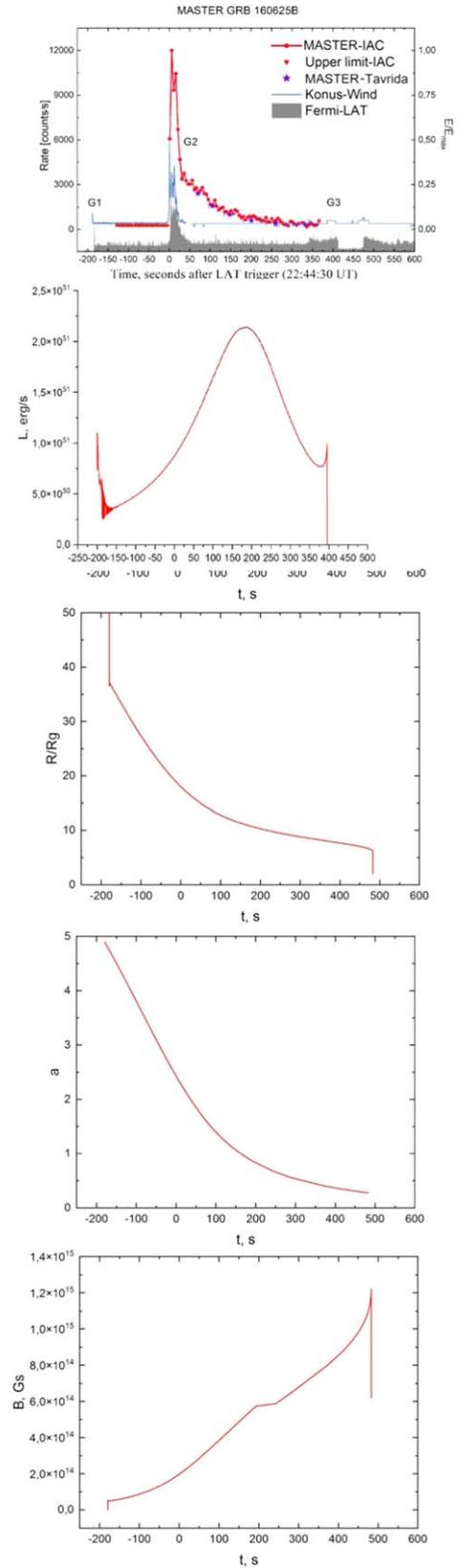

**Figure 11.** Diagram of the GRB 160625B event in the observer's frame of reference (MASTER photometry is in Table 4 in the Appendix) is shown. In the top diagram, we have shown three principal G1 events—the position of the precursor ($\Delta t = -200$ s), the main peak of the G2 pulse ($t = 0$), and finally the last episode of increased G3 activity ($\Delta t = 400$–500 s).

Obviously, the main time intervals between the events of the GRB G1, G2, and G3 are in good agreement with observations. Of course, we do not pretend to describe a detailed light curve within the framework of our approximate model. We can only compare with observations the total energy release of the central engine at times G1, G2, and G3.

We associate a slight increase in the gamma flux in G3 with an increase in the power of the central engine at the moment when the spinar loses its stability and collapses into a black hole. Note that such a three-stage collapse occurs only when the mass of the collapsing nucleus exceeds but is comparable to the Oppenheimer–Volkov limit.

## 6. Discussion

We interpret the features of the GRB 160625B radiation within the framework of the model of a three-stage magneto-rotational collapse of the core of a massive star whose mass exceeds but is comparable to the Oppenheimer–Volkov limit. In this model, the first G1 event in the spinar paradigm will be explained by an abrupt stop of the nucleus collapse at the centrifugal barrier and the formation of a spinar. In this case, half of the accumulated gravitational energy is converted into rotational energy, and the second half can be converted into jet energy along the spinar's axis of rotation. Further, as the rotational moment is lost, the spinar is compressing, gradually increasing its rotational energy, mainly emitted along the axis of rotation with power proportional to the Umov–Poynting vector electromagnetic energy flow $E_{G1} = U_m \omega$. Of course, the observer will not see a smooth curve, since the jet must accumulate enough energy to pierce the progenitor shell (Lipunov & Gorbovskoy 2007). Naturally, in a model based on the conservation laws with an approximate description of the torque dissipation, one should not rely on the exact repetition of the light curve at the moment of the main pulse G2. Therefore, we only achieved the coincidence of the moment of the maximum energy release with observations and the interval preceding the collapse of a heavy spinar with a powerful magnetic field and a mass exceeding the Oppenheimer–Volkov limit for a nonrotating neutron star.

However, the correspondence between the conclusions of this model to the results of observations remains to be observed.

In this paper, we presented multiwavelength observations of GRB 160625B, one of the brightest GRBs in the history of their study. The authors of the article have already published a paper, concerning the first discovery of variable polarization of the intrinsic emission of GRBs (Troja et al. 2017). Here, we have concentrated on the temporal behavior of its intrinsic electromagnetic radiation. This is the first time we publish optical and gamma-ray data, recorded during the time of the operation of the central engine of GRB 160625B. We made an attempt to find traces of the duality of the GRB system. For the first time, we have suspected traces of the dual nature of long GRBs. Of course, we do not have 100% proof of this scenario, but this study can serve as an example of searching for and finding the dual nature of long GRBs.

MASTER and the Lomonosov Space Observatory are supported by Lomonosov Moscow State University Development program (equipment). V.L. and P.B. are supported by RFBR grant 19-29-11011. N.B. is supported by FZZE-2020-0017 and used the UNU "Astrophysical Complex of MSU-ISU" (agreement EB-075-15-2021-675). We are grateful to the reviewer for the valuable comments.

## Appendix

In this appendix we present the full photometry of GRB 160625B optical counterpart by MASTER (Table 4) and the explanation for the Abbe test (Figure 12) for quasiperiodic optical variavbility of the intrinsic optical radiation GRB 160625B, detected by MASTER (see Section 3).

Table 4
GRB 160625B optical counterpart photometry and unfiltered Magnitude Limits of MASTER

| JD of Beginning of Exposure | T(mid)-T(GBM) [s] | T(mid)-T(LAT) [s] | Limit/ Mag | Error of Magnitude | Exp. Time | Telescope |
| --- | --- | --- | --- | --- | --- | --- |
| 2457565,44529209 | 59.4570155 | −129.0833418 | <12.0 | | 5 | MASTER-IAC VWF (East) |
| 2457565,44534996 | 64.45698838 | −124.083369 | <12.0 | | 5 | MASTER-IAC VWF (East) |
| 2457565,44540783 | 69.45696126 | −119.0833961 | <12.0 | | 5 | MASTER-IAC VWF (East) |
| 2457565,44546570 | 74.45693414 | −114.0834232 | <12.0 | | 5 | MASTER-IAC VWF (East) |
| 2457565,44552357 | 79.45686679 | −109.0834906 | <12.0 | | 5 | MASTER-IAC VWF (East) |
| 2457565,44558145 | 84.4577248 | −104.0826326 | <12.0 | | 5 | MASTER-IAC VWF (East) |
| 2457565,44563932 | 89.45769768 | −99.08265967 | <12.0 | | 5 | MASTER-IAC VWF (East) |
| 2457565,44569719 | 94.45767056 | −94.08268679 | <12.0 | | 5 | MASTER-IAC VWF (East) |
| 2457565,44575506 | 99.4576032 | −89.08275415 | <12.0 | | 5 | MASTER-IAC VWF (East) |
| 2457565,44581293 | 104.4575761 | −84.08278127 | <12.0 | | 5 | MASTER-IAC VWF (East) |
| 2457565,44587081 | 109.4584341 | −79.08192326 | <12.0 | | 5 | MASTER-IAC VWF (East) |
| 2457565,44592868 | 114.458407 | −74.08195038 | <12.0 | | 5 | MASTER-IAC VWF (East) |
| 2457565,44598655 | 119.4583396 | −69.08201773 | <12.0 | | 5 | MASTER-IAC VWF (East) |
| 2457565,44604442 | 124.4583125 | −64.08204485 | <12.0 | | 5 | MASTER-IAC VWF (East) |
| 2457565,44610229 | 129.4582854 | −59.08207197 | <12.0 | | 5 | MASTER-IAC VWF (East) |
| 2457565,44616016 | 134.4582583 | −54.08209909 | <12.0 | | 5 | MASTER-IAC VWF (East) |
| 2457565,44621803 | 139.4582311 | −49.08212621 | <12.0 | | 5 | MASTER-IAC VWF (East) |
| 2457565,44627590 | 144.458204 | −44.08215333 | <12.0 | | 5 | MASTER-IAC VWF (East) |
| 2457565,44633377 | 149.4581769 | −39.08218045 | <12.0 | | 5 | MASTER-IAC VWF (East) |
| 2457565,44639164 | 154.4581498 | −34.,08220757 | <12.0 | | 5 | MASTER-IAC VWF (East) |
| 2457565,44644951 | 159.4580824 | −29.08227493 | <12.0 | | 5 | MASTER-IAC VWF (East) |
| 2457565,44650738 | 164.4580553 | −24.08230205 | <12.0 | | 5 | MASTER-IAC VWF (East) |
| 2457565,44656546 | 169.4761733 | −19.06418402 | <12.0 | | 5 | MASTER-IAC VWF (East) |
| 2457565,44662313 | 174.4588862 | −14.08147116 | <12.0 | | 5 | MASTER-IAC VWF (East) |
| 2457565,44668100 | 179.4588188 | −9.08153851 | <12.0 | | 5 | MASTER-IAC VWF (East) |
| 2457565,44673887 | 184.4587917 | −4.08156563 | <12.0 | | 5 | MASTER-IAC VWF (East) |
| 2457565,44679674 | 189.4587646 | 0.918407249 | 8.604 | 0.051 | 5 | MASTER-IAC VWF (East) |
| 2457565,44685461 | 194.4587375 | 5.918380129 | 7.864 | 0.048 | 5 | MASTER-IAC VWF (East) |
| 2457565,44691248 | 199.4587104 | 10.91835301 | 8.138 | 0.049 | 5 | MASTER-IAC VWF (East) |
| 2457565,44697035 | 204.4586832 | 15.91832589 | 8.014 | 0.049 | 5 | MASTER-IAC VWF (East) |
| 2457565,44702822 | 209.4586561 | 20.91829877 | 8.497 | 0.051 | 5 | MASTER-IAC VWF (East) |
| 2457565,44708609 | 214.458629 | 25.91827165 | 8.890 | 0.053 | 5 | MASTER-IAC VWF (East) |
| 2457565,44714396 | 219.4585616 | 30.9182043 | 9.244 | 0.055 | 5 | MASTER-IAC VWF (East) |
| 2457565,44720183 | 224.4585345 | 35.91817718 | 9,126 | 0,054 | 5 | MASTER-IAC VWF (East) |
| 2457565,44725970 | 229.4585074 | 40.91815006 | 9.279 | 0.055 | 5 | MASTER-IAC VWF (East) |
| 2457565,44731757 | 234.4584803 | 45.91812294 | 9.356 | 0.085 | 5 | MASTER-IAC VWF (East) |
| 2457565,44737544 | 239.4584532 | 50.91809582 | 9.356 | 0.085 | 5 | MASTER-IAC VWF (East) |
| 2457565,44743331 | 244.458426 | 55.9180687 | 9.270 | 0.055 | 5 | MASTER-IAC VWF (East) |
| 2457565,44749119 | 249.4592438 | 60.91888647 | 9,506 | 0,083 | 5 | MASTER-IAC VWF (East) |

Table 4
(Continued)

| JD of Beginning of Exposure | T(mid)-T(GBM) [s] | T(mid)-T(LAT) [s] | Limit/ Mag | Error of Magnitude | Exp. Time | Telescope |
| --- | --- | --- | --- | --- | --- | --- |
| 2457565,44754906 | 254.4592167 | 65.91885935 | 9.452 | 0.083 | 5 | MASTER-IAC VWF (East) |
| 2457565,44757008 | 258.7749261 | 70.23456872 | 9.611 | 0.036 | 10 | MASTER- Tavrida |
| 2457565,44760693 | 259.4591896 | 70.91883223 | 9.593 | 0.083 | 5 | MASTER-IAC VWF (East) |
| 2457565,44766480 | 264.4591625 | 75.91880511 | 9.521 | 0.083 | 5 | MASTER-IAC VWF (East) |
| 2457565,44772268 | 269.4599802 | 80.91962289 | 9.446 | 0.083 | 5 | MASTER-IAC VWF (East) |
| 2457565,44778055 | 274.4599531 | 85.91959577 | 9.607 | 0.083 | 5 | MASTER-IAC VWF (East) |
| 2457565,44783842 | 279.459926 | 90.91956865 | 9.638 | 0.084 | 5 | MASTER-IAC VWF (East) |
| 2457565,44789629 | 284.4598989 | 95.91954153 | 9.814 | 0.090 | 5 | MASTER-IAC VWF (East) |
| 2457565,44795416 | 289.4598718 | 100.9195144 | 10.006 | 0.102 | 5 | MASTER-IAC VWF (East) |
| 2457565,44791100 | 293.230409 | 104.6900517 | 9.801 | 0.024 | 20 | MASTER-IAC (East) |
| 2457565,44791100 | 293.230409 | 104.6900517 | 9.905 | 0.032 | 20 | MASTER-IAC (West) |
| 2457565,44801203 | 294.4598446 | 105.9194873 | 10.007 | 0.102 | 5 | MASTER-IAC VWF (East) |
| 2457565,44792707 | 294.6188545 | 106.0784972 | 10.083 | 0.027 | 20 | MASTER- Tavrida |
| 2457565,44806990 | 299.4597773 | 110.9194199 | 9.838 | 0.091 | 5 | MASTER-IAC VWF (East) |
| 2457565,44812777 | 304.4597502 | 115.9193928 | 10.092 | 0.110 | 5 | MASTER-IAC VWF (East) |
| 2457565,44818564 | 309.459723 | 120.9193657 | 10.273 | 0.129 | 5 | MASTER-IAC VWF (East) |
| 2457565,44824351 | 314.4596959 | 125.9193386 | 10.185 | 0.119 | 5 | MASTER-IAC VWF (East) |
| 2457565,44830138 | 319.4596688 | 130.9193115 | 10.134 | 0.114 | 5 | MASTER-IAC VWF (East) |
| 2457565,44835925 | 324.4596417 | 135.9192843 | 10.388 | 0,145 | 5 | MASTER-IAC VWF (East) |
| 2457565,44841713 | 329.4604595 | 140.9201021 | 10.374 | 0,143 | 5 | MASTER-IAC VWF (East) |
| 2457565,44847500 | 334.4604323 | 145.920075 | 10,517 | 0,165 | 5 | MASTER-IAC VWF (East) |
| 2457565,44835424 | 336.526359 | 147.9860017 | 10.570 | 0,020 | 30 | MASTER- Tavrida |
| 2457565,44853287 | 339.4604052 | 150.9200479 | 10.331 | 0.137 | 5 | MASTER-IAC VWF (East) |
| 2457565,44859074 | 344.4603781 | 155.9200207 | 10.504 | 0.163 | 5 | MASTER-IAC VWF (East) |
| 2457565,44847700 | 347.1328193 | 158.5924619 | 10.542 | 0.048 | 30 | MASTER-IAC (East) |
| 2457565,44847700 | 347.1328193 | 158.5924619 | 10.673 | 0.045 | 30 | MASTER-IAC (West) |
| 2457565,44864861 | 349.460351 | 160.9199936 | 10.270 | 0.129 | 5 | MASTER-IAC VWF (East) |
| 2457565,44870648 | 354.4603239 | 165.9199665 | 10.354 | 0.140 | 5 | MASTER-IAC VWF (East) |
| 2457565,44876435 | 359.4602565 | 170.9198992 | 10.629 | 0.185 | 5 | MASTER-IAC VWF (East) |
| 2457565,44882225 | 364.4628445 | 175.9224872 | 10.828 | 0.225 | 5 | MASTER-IAC VWF (East) |
| 2457565,44888009 | 369.4602023 | 180.9198449 | 10.720 | 0.202 | 5 | MASTER-IAC VWF (East) |
| 2457565,44893796 | 374.4601751 | 185.9198178 | 11.043 | 0.276 | 5 | MASTER-IAC VWF (East) |
| 2457565,44899584 | 379.4609929 | 190.9206356 | 10.779 | 0.215 | 5 | MASTER-IAC VWF (East) |
| 2457565,44905371 | 384.4609658 | 195.9206084 | 10.705 | 0.199 | 5 | MASTER-IAC VWF (East) |
| 2457565,44890168 | 388.825157 | 200.2847996 | 11.099 | 0.015 | 40 | MASTER-Tavrida |
| 2457565,44911158 | 389.4609387 | 200.9205813 | 10.746 | 0.208 | 5 | MASTER-IAC VWF (East) |
| 2457565,44896500 | 394.2960185 | 205.7556611 | 11.050 | 0.068 | 40 | MASTER-IAC (East) |
| 2457565,44896500 | 394.2960185 | 205.7556611 | 11.163 | 0.050 | 40 | MASTER-IAC (West) |
| 2457565,44916945 | 394.4609116 | 205.9205542 | 10.663 | 0.191 | 5 | MASTER-IAC VWF (East) |

Table 4
(Continued)

| JD of Beginning of Exposure | T(mid)-T(GBM) [s] | T(mid)-T(LAT) [s] | Limit/ Mag | Error of Magnitude | Exp. Time | Telescope |
|---|---|---|---|---|---|---|
| 2457565,44922732 | 399.4608844 | 210.9205271 | 10.804 | 0.220 | 5 | MASTER-IAC VWF (East) |
| 2457565,44928519 | 404.4608573 | 215.9205 | 10.985 | 0.262 | 5 | MASTER-IAC VWF (East) |
| 2457565,44934306 | 409,4608302 | 220,9204728 | 10.938 | 0.250 | 5 | MASTER-IAC VWF (East) |
| 2457565,44940094 | 414,461648 | 225,9212906 | 11.164 | 0.308 | 5 | MASTER-IAC VWF (East) |
| 2457565,44945881 | 419,4616209 | 230,9212635 | 11.139 | 0.301 | 5 | MASTER-IAC VWF (East) |
| 2457565,44951668 | 424.4615937 | 235.9212364 | 10.820 | 0.223 | 5 | MASTER-IAC VWF (East) |
| 2457565,44957455 | 429.4615666 | 240.9212093 | 11.302 | 0.348 | 5 | MASTER-IAC VWF (East) |
| 2457565,44963242 | 434.4615395 | 245.9211821 | 11.291 | 0.345 | 5 | MASTER-IAC VWF (East) |
| 2457565,44969029 | 439.4614721 | 250.9211148 | 10.916 | 0.245 | 5 | MASTER-IAC VWF (East) |
| 2457565,44974816 | 444.461445 | 255.9210877 | 11.230 | 0.327 | 5 | MASTER-IAC VWF (East) |
| 2457565,44954363 | 449.2896318 | 260.7492745 | 11.561 | 0.013 | 50 | MASTER- Tavrida |
| 2457565,44980603 | 449.4614179 | 260.9210606 | 11.131 | 0.299 | 5 | MASTER-IAC VWF (East) |
| 2457565,44986390 | 454.4613908 | 265.9210334 | 10.877 | 0.236 | 5 | MASTER-IAC VWF (East) |
| 2457565,44992177 | 459.4613637 | 270.9210063 | 11.137 | 0.301 | 5 | MASTER-IAC VWF (East) |
| 2457565,44997964 | 464.4613365 | 275.9209792 | 11.428 | 0.387 | 5 | MASTER-IAC VWF (East) |
| 2457565,45003751 | 469.4613094 | 280.9209521 | 11.770 | 0.506 | 5 | MASTER-IAC VWF (East) |
| 2457565,45009538 | 474.4612823 | 285.9209249 | 11.748 | 0.498 | 5 | MASTER-IAC VWF (East) |
| 2457565,45015325 | 479.4612552 | 290.9208978 | 12.091 | 0.636 | 5 | MASTER-IAC VWF (East) |
| 2457565,45021112 | 484.4611878 | 295.9208305 | 11.243 | 0.331 | 5 | MASTER-IAC VWF (East) |
| 2457565,45026899 | 489.4611607 | 300.9208034 | 11.271 | 0.339 | 5 | MASTER-IAC VWF (East) |
| 2457565,45032687 | 494.4620187 | 305.9216614 | 11.624 | 0.453 | 5 | MASTER-IAC VWF (East) |
| 2457565,45038475 | 499.4628365 | 310.9224791 | 11.917 | 0.563 | 5 | MASTER-IAC VWF (East) |
| 2457565,45044262 | 504.4628094 | 315.922452 | 11.549 | 0.427 | 5 | MASTER-IAC VWF (East) |
| 2457565,45055836 | 514.4627551 | 325.9223978 | 11.422 | 0.385 | 5 | MASTER-IAC VWF (East) |
| 2457565,45028300 | 518.1712005 | 329.6308431 | 11.934 | 0.047 | 60 | MASTER-IAC (East) |
| 2457565,45028300 | 518.1712005 | 329.6308431 | 12.117 | 0.037 | 60 | MASTER-IAC (West) |
| 2457565,45061623 | 519.4626878 | 330.9223304 | 11.584 | 0.439 | 5 | MASTER-IAC VWF (East) |
| 2457565,45030880 | 520.4003575 | 331.8600002 | 11.720 | 0.013 | 60 | MASTER-Tavrida |
| 2457565,45067410 | 524.4626607 | 335.9223033 | 12.474 | 0.814 | 5 | MASTER-IAC VWF (East) |
| 2457565,45073197 | 529.4626335 | 340.9222762 | 11.887 | 0.552 | 5 | MASTER-IAC VWF (East) |
| 2457565,45078984 | 534.4626064 | 345.9222491 | 11.443 | 0.392 | 5 | MASTER-IAC VWF (East) |
| 2457565,45084771 | 539.4625793 | 350.9222219 | 11.827 | 0.528 | 5 | MASTER-IAC VWF (East) |
| 2457565,45090558 | 544.4625522 | 355.9221948 | 11.997 | 0.596 | 5 | MASTER-IAC VWF (East) |
| 2457565,45096345 | 549.4625251 | 360.9221677 | 11.799 | 0.517 | 5 | MASTER-IAC VWF (East) |
| 2457565,45102132 | 554.4624979 | 365.9221406 | 11.158 | 0.307 | 5 | MASTER-IAC VWF (East) |

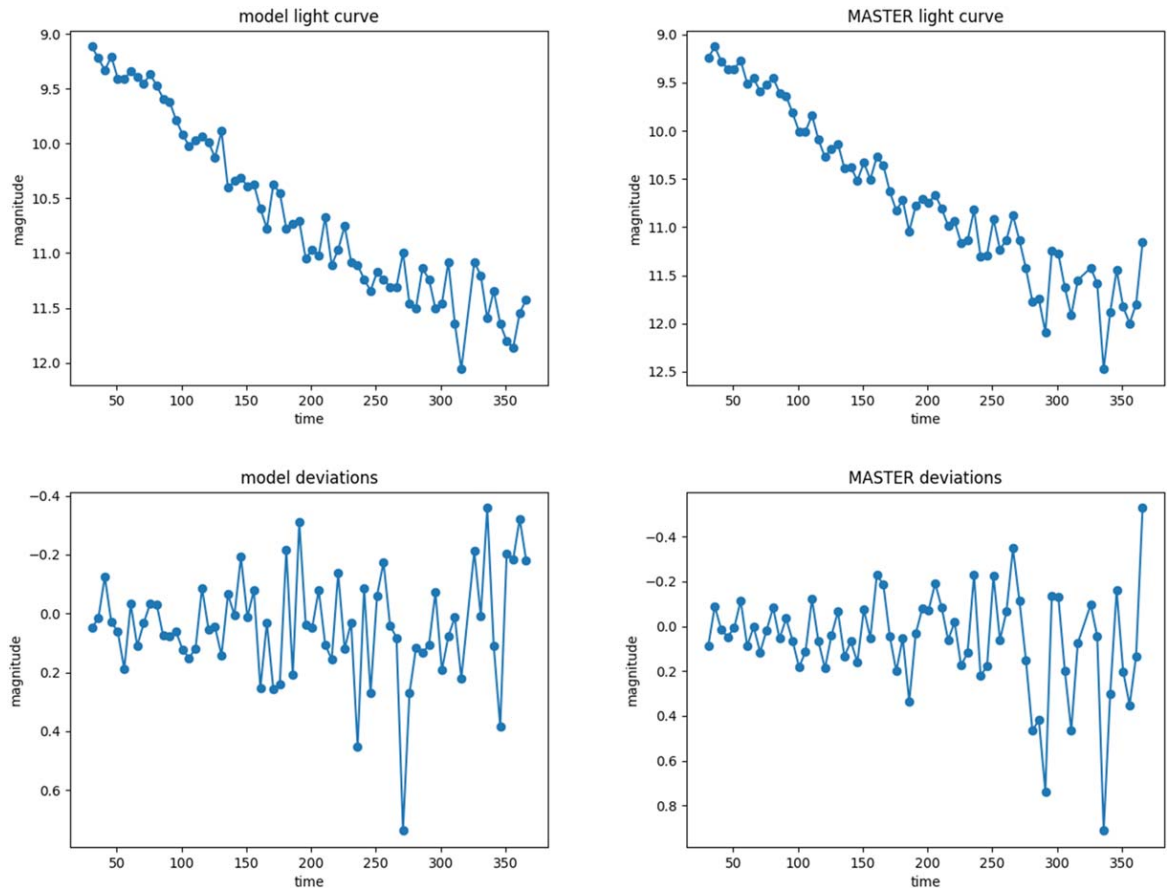

**Figure 12.** The Abbe test, satisfied with a significance level of $\alpha = 0.01$, shows that for a random normally distributed noise, the probability of obtaining a gamma coefficient satisfying the observed condition is 1%. By replacing the noise in the Abbe test with Poisson noise and numerically modeling the situation, we calculated the probability to satisfy the inequality of the Abbe test for a completely Poisson random noise. This calculation was carried out by the Monte Carlo method. A comparison between one of the model curves and the observed curve is presented here. Having generated 1 million model curves, we can say that the probability of Poisson noise satisfying the Abbe criterion is 0.023, i.e., $\alpha = 0.023$. Thus, this light curve has a deterministic component with a probability of 97.7%. But the Abbe test does not state anything about the period of this component. We estimated the value of the possible period by the peaks of the periodogram

## ORCID iDs

V. M. Lipunov ● https://orcid.org/0000-0003-3668-1314
G. V. Lipunova ● https://orcid.org/0000-0003-4515-8955
D. Frederiks ● https://orcid.org/0000-0002-1153-6340
R. Rebolo ● https://orcid.org/0000-0003-3767-7085
A. F. Iyudin ● https://orcid.org/0000-0001-9318-1258
A. Gabovich ● https://orcid.org/0000-0002-8194-6484
A. Tsvetkova ● https://orcid.org/0000-0003-0292-6221
N. M. Budnev ● https://orcid.org/0000-0002-2104-6687
O. A. Gress ● https://orcid.org/0000-0001-7960-7582
I. Gorbunov ● https://orcid.org/0000-0003-2380-141X
V. Vladimirov ● https://orcid.org/0000-0001-7957-2156
C. Francile ● https://orcid.org/0000-0002-9148-1780